\documentclass[aps,showpacs,floatfix]{revtex4}
\usepackage{graphicx}

\makeatletter
\newbox\slashbox \setbox\slashbox=\hbox{\large$/$}
\def\pslash#1{\setbox\@tempboxa=\hbox{$#1$}
  \@tempdima=0.5\wd\slashbox \advance\@tempdima 0.5\wd\@tempboxa
  \copy\slashbox \kern-\@tempdima \box\@tempboxa}
\def\slash{\protect\pslash}
\makeatother

\newcommand{\mat}{\left ( \begin{array}{cc}}
\newcommand{\emat}{\end{array} \right )}
\newcommand{\be}{\begin{eqnarray}}
\newcommand{\ee}{\end{eqnarray}}
\newcommand{\ba}{\begin{array}}
\newcommand{\ea}{\end{array}}

\def\SB{S$\chi$SB}
\def\fm{\,\mathrm{fm}}
\def\beq{\begin{equation}}
\def\eeq{\end{equation}}

\begin{document}

\title{Chiral phase transition and Anderson localization\\
 in the Instanton Liquid Model for QCD}
\author{Antonio M. Garc\'{\i}a-Garc\'{\i}a}
\affiliation{Physics Department, Princeton University, Princeton,
New Jersey 08544, USA}
\affiliation{The Abdus Salam International Centre for Theoretical
Physics, P.O.B. 586, 34100 Trieste, Italy}
\author{James C. Osborn}
\affiliation{Physics Department \& Center for Computational Science,
 Boston University, Boston, MA 02215, USA}

\begin{abstract}
We study the spectrum and eigenmodes of the
 QCD Dirac operator in a gauge background given by an Instanton Liquid Model
 (ILM) at temperatures around the chiral phase transition.
Generically we find the Dirac eigenvectors
become more localized as the temperature is increased.
At the chiral phase transition, both the 
 low lying eigenmodes and the spectrum of the QCD Dirac operator 
 undergo a transition to localization similar 
 to the one observed in a disordered conductor. This 
suggests that Anderson localization is the fundamental
mechanism driving the chiral phase transition.
We also find an additional temperature dependent mobility 
 edge (separating delocalized from localized eigenstates) in the 
 bulk of the spectrum which moves toward lower eigenvalues
as the temperature is increased.
In both regions, the origin and the bulk,
the transition to localization exhibits features of a 3D Anderson transition
including multifractal eigenstates and spectral
properties that are well described by critical statistics.
Similar results are obtained in both the quenched and the unquenched
case though the critical temperature in the unquenched case is lower.
Finally we argue that our findings are not in principle restricted to the ILM
 approximation and may also be found in lattice simulations.
\end{abstract}
\maketitle

\section{Introduction} 

By now it is well established that many of the properties of light hadrons
are dictated by spontaneous chiral symmetry breaking
(\SB) due to the nonperturbative structure of the QCD vacuum.
However, in the high temperature limit, non perturbative
effects are suppressed and chiral symmetry is
  eventually restored at a certain critical temperature.
Recent lattice simulations suggest furthermore that both the
restoration of chiral symmetry and the confinement-deconfinement
transition occur at the same temperature \cite{lattrans}.
A detailed account of the microscopic mechanisms leading to
\SB{} and confinement and the eventual restoration and deconfinement
transition remains among the most formidable challenges in hadronic
physics.
In recent years the combination of experimental results and
novel lattice techniques have greatly clarified the nature and
properties of these transitions.
However we are still far from a full microscopic understanding of the
chiral phase transition and its relation to the
confinement-deconfinement transition.
In this paper we examine how the phenomenon of Anderson
localization \cite{anderson}, originally introduced in the field of
condensed matter physics, may help to understand the mechanism
driving the chiral phase transition.
Specifically, in the context of an Instanton Liquid Model (ILM),
we show that near the temperature around which the chiral 
phase transition occurs the low lying eigenmodes
 of the QCD Dirac operator undergo
 a localization-delocalization transition
consistent with an Anderson transition (AT)  
 characterized by 
 multifractal eigenstates and critical statistics \cite{anderson}.
In order to make accessible our findings to a broader audience, we
first provide short summaries on the role of instantons in the
non-perturbative structure of QCD,
similarities between the QCD vacuum and a disordered medium, the QCD chiral
phase transition and finally a brief overview on the properties of a
disordered conductor at an AT.
In section \ref{secResults} we will present our results for the ILM simulations
in both the quenched case and the unquenched case for two and three flavors.
In section \ref{secDiscuss} we discuss the relationship of our results with
current lattice simulations followed by conclusions in section \ref{secConcl}.

\subsection{Instanton Liquid Models and \SB}

Instantons \cite{polyakov} are
classical solutions of the Yang-Mills equations of motion which minimize the
action in Euclidean space. They are believed to be the leading
semiclassical contribution to the bosonic part of the
QCD path integral.  However the
construction of a consistent QCD vacuum based on instantons faces
serious technical difficulties.  Exact multi-instanton solutions are
hard to obtain since the Yang-Mills equations of motion for QCD are
nonlinear and therefore a superposition of single instanton
contributions is not itself a solution.  Additionally quantum
corrections may spoil the semiclassical picture implicitly assumed of
a QCD vacuum composed of instantons well separated and weakly
interacting.  These problems have been overcome either by invoking
variational principles \cite{diakonov} or by phenomenologically fixing
certain parameters of the instanton ensemble.  The latter case,
usually referred to as the instanton liquid model (ILM)
\cite{shuryak} (for a modern review see \cite{SS97}),
 yields accurate estimates of vacuum condensates and
hadronic correlation functions \cite{SSV} for most light hadrons just
by setting the mean distance between instantons to be ${\bar R}\approx 1
\fm$ (which corresponds to a density $N/V \sim 1\fm^{-4}$) 
 and the mean instanton size ${\bar \rho} \approx 1/3 \fm$.
Lattice simulations have also supported the picture
 of a QCD vacuum dominated by instantons \cite{latins}.

One of the main success of the
ILM is its convincing explanation of how chiral symmetry 
 is spontaneously broken in the QCD vacuum \cite{shuryak,diakonov}.
 The starting point is the fact 
that the Euclidean QCD Dirac operator has an
 exact zero eigenvalue in the field of an instanton.
In an ensemble of (anti-)instantons  
 the overlap of neighboring instantons
makes the  would-be zero modes effectively split around zero. 
In the limit of large number of (anti-)instantons a continuous band spectrum 
is formed with a spectral density finite at zero. 
The spectral properties of the low lying modes of the Dirac
 operator are thus controlled by these 
 non-perturbative configurations.
It turns out that the order parameter 
associated with spontaneous chiral symmetry breaking
  $\langle {\bar \psi} \psi \rangle$  (the chiral condensate)
 is indeed related to the spectral 
density $\rho(\epsilon)$ of the QCD Dirac operator around zero
 through the Banks-Casher
 relation \cite{bank},
\be
\langle {\bar \psi} \psi \rangle =
 - \lim_{\epsilon \to 0} \lim_{V \to \infty} \frac{\pi \rho(\epsilon)}{V} ~,
\ee 
where $V$ is the space-time volume.
We are therefore interested in modeling the lowest eigenvalues of the
Dirac operator.
The ILM provides a phenomenological model for these low energy modes of the
QCD Dirac operator.
In a basis of $N/2$ instantonic zero modes $\psi_{I}^{(i)}$
and $N/2$ anti-instantonic zero modes $\psi_{A}^{(j)}$ the matrix
elements of the Dirac operator take the form (for zero quark mass)
\be
\label{dilm}
{\cal D}_{ILM} =
 \left (\begin{array}{cc} 0 & iT_{IA}\\
iT_{IA}^\dagger & 0 \end{array} \right) ~.
\ee
In the limit of large separations ${\vec R}_{ij} = {\vec z}_i - {\vec z}_j$
between the center of instanton $i$ and anti-instanton $j$ the
elements of the $N/2 \times N/2$ overlap matrix $T_{IA}$ 
decay as a power-law,
\be
 \left[ T_{IA} \right]_{i,j} ~=~
 \langle\psi_{I}^{(i)}|\,{\slash D}\,|\psi_{A}^{(j)}\rangle
 ~\sim~ \rho_i  \rho_j  /  |{\vec R}_{ij}|^3
\label{olpl}
\ee
 with $\rho_i$ and $\rho_j$ their respective sizes
 and ${\slash D}$ is the Dirac operator.
Many properties of the QCD vacuum can then be estimated
from the matrix elements of $T_{IA}$.
For example the chiral condensate can be calculated as
$\langle {\bar \psi} \psi \rangle \approx -(215\mathrm{MeV})^3$
\cite{diakonov,SS97}
which is reasonably close to the experimental value.   

\subsection{The QCD vacuum as a disordered medium}

In recent years it has been
 suggested \cite{diakonov1,smilga,zahed,VO} that the
\SB{} in QCD and  the phenomenon of conductivity in a disordered medium 
 have similar physical origins.
Conductivity in a disordered sample is produced by electrons that although
initially bound to an impurity may get delocalized by orbital overlapping with nearby impurities. 
Similarly, in the QCD vacuum, the zero modes of the Dirac operator though 
initially bound to an instanton may get delocalized due to  
the strong overlap with other instantons. As a consequence, the
 chiral condensate becomes nonzero and chiral symmetry is broken.

In the case of electrons the overlap is effective only among nearest neighbors
since the electron wavefunction has an exponential tail. 
On the contrary, in the QCD vacuum, quark zero modes can travel long distances
in the QCD vacuum.
Hopping is not restricted to nearest neighbors since the probability for a
quark to hop between an instanton and an anti-instanton
 has a power-law tail in the pseudo-particle distance (equation \ref{olpl}).
As we discuss next, eigenvectors
 of quantum systems with this kind of long range 
 hopping are delocalized in four (and higher) dimensions \cite{levitov90,par}.

Localization properties of a disordered system are
strongly influenced by factors such as the dimensionality of the space, 
 the strength of disorder and the  
 hopping range. 
For short range hopping  
it is well known that,
in three and higher dimensions and for a certain range
 of impurity amount, a system will have a mobility edge 
 separating localized from a delocalized eigenstates.  
Eigenstates around the mobility edge (the Anderson transition region) 
are multifractal and the level statistics (commonly referred to as 
\emph{critical statistics} \cite{kravtsov97})
 are universal and intermediate between Wigner-Dyson (WD) and Poisson statistics
(for a review see \cite{janssen}).

An Anderson transition is also observed in disordered systems with
 power-law hopping \cite{ono,levitov90,par} provided that the exponent of the 
 hopping decay matches the dimension of the space.
If the exponent is larger (smaller) than the dimension of the space then
 the eigenstates becomes localized (delocalized).
In this case the localization properties in the thermodynamic limit 
 depend only on the power-law decay and not on the strength of disorder. 
Since in QCD at zero temperature the dimensionality of the space (four)
 is larger than the hoping decay power (three, see equation \ref{olpl}),
 one expects the 
 eigenstates to be delocalized independently of the density of instantons
(analogous to the number of impurities which determines the disorder strength).
Indeed in a recent paper \cite{aj} we showed that an effective random matrix
 model incorporating the phenomenological power-law decay for the
 matrix elements of the QCD Dirac operator in a basis of zero modes  
 describes the spectral correlations of the QCD Dirac operator 
 well beyond the Thouless energy $ E_c \sim 
 {F_{\pi}^{2}}/(|\langle {\bar \psi} \psi \rangle| L^2)$
 ($F_{\pi}$ is the pion decay constant and $L$ the length of the system)
 which sets the limit of applicability of
 standard chiral random matrix theory \cite{VO}.
However at high temperatures, QCD becomes effectively three dimensional,
 the chiral symmetry is restored and
 we expect to see localized states appear in the spectrum.
One of the main aims of this paper is precisely to 
 investigate the relationship between the vanishing of the
 chiral condensate and the localization of the eigenstates of the
 QCD Dirac operator in a background of instantons.

\subsection{ILM at finite temperature:
            chiral phase transition and localization}

In this section we summarize how the ILM is extended to finite temperature
 and also to what extent reproduces the main
 features of QCD at temperatures around the chiral phase transition.
It is clear that for sufficiently high temperature (beyond the chiral phase
 transition) QCD becomes perturbative and instanton configurations should be
 exponentially suppressed.
However at temperatures around and below the chiral phase transition the
 density of instantons is still large and it is expected that they play an
 important role in the description of the phase transition.
The generalization of the ILM to finite temperature
 was presented in \cite{SS97}.
Below we provide a brief summary of its main features and refer to 
\cite{SS97} for details:
\begin{itemize}
\item
 The effective coupling constant $g(T)$ decreases (QCD is an
 asymptotically free theory) as temperature increases. The density of
 instantons $\rho$ decreases with temperature since $\rho \propto
 \exp(-S)$ with $S=8\pi^2/g^2(T)$.  At the restoration temperature the
 density is still roughly half of the zero temperature value.
\item
 Instanton like solutions (usually called calorons) are still zero modes of
 the QCD Dirac operator at finite temperature. 
 However the decay of the associated fermionic zero mode $\psi(R)
 \sim e^{-\pi T R}$ ($R$ is the 3D instanton--anti-instanton distance)
 is exponential (instead of power-law) in the spatial dimensions
 and oscillatory in the Euclidean time dimension.
\item
 The probability $|T_{IA}| \sim e^{-\alpha T R}$ ($\alpha$ a numerical factor
 of order unity) for a fermionic zero mode to hop from an instanton to an
 anti-instanton has an exponential tail in the spatial dimensions thus
 restricting the overlap only to nearby instantons.
 On the contrary, in the time direction the probability is oscillatory.
 This suggests that at finite temperature the isolated fermionic zero modes
 are exponentially
 localized in space but delocalized in time, thus for localization studies
 the dimension of the system is three and not four. In this case,
 since the effective hopping is short range, the global localization
 properties of the eigenstates of QCD Dirac operator are expected to
 depend strongly on the density of instantons as for normal disordered
 conductors in 3D.
\item
 The exponentially small tail of the overlap matrix elements
 together with the oscillatory behavior in Euclidean time suggest that in
 the high temperature limit (higher than the restoration 
 temperature) the instantons and anti-instantons are
 paired in molecules aligned along the time direction.
 As a consequence the spectral density of the QCD Dirac operator develops a
 gap near zero
 making, according the Banks-Casher relation, the chiral condensate vanish. 
\end{itemize}
Although certainly appealing, the validity of the ILM at
 finite temperature as an accurate model for QCD around the phase
 transition should be corroborated by lattice simulations.
Recent lattice studies \cite{lattice} suggest that this is the case though
 further investigations are needed to fully settle this issue.

\subsection{Anderson transition: description and signatures
\label{secpq}}

A disordered system with short range hopping in three or more
 dimensions displays a metal-insulator transition or Anderson
 transition (AT) around the band center for a critical amount of disorder.
By critical amount of disorder we mean if the disorder is
 increased all the eigenstates become exponentially localized.
For a disorder strength below the critical one, the system has
a mobility edge at a certain energy which separates localized  
from delocalized states.
Its position moves away from the band center as the disorder is decreased.
Delocalized eigenstates, typical of a metal, are extended through the sample 
 and the level statistics agree with the random matrix prediction for the
 appropriate symmetry.
In this paper we will focus on QCD with three colors which corresponds to
 random matrix predictions from the Gaussian Unitary Ensemble (GUE)
 \cite{mehta}.
The level statistics obtained from random matrix theory are also referred
to as Wigner-Dyson statistics (WD).
On the contrary, localized eigenstates, typical of an insulator,
are characterized by an 
exponential decay $|\psi(r)| \sim e^{-r/\xi}$ with the 
 localization length $\xi$ smaller than the system size.
The spectral correlations in this case 
 are well described by Poisson statistics \cite{mehta} corresponding to
 uncorrelated eigenvalues.

Around the mobility edge the system is referred to as being at the AT 
(or at the metal-insulator transition). Signatures of
this transition are found in both the spectrum and the eigenfunctions.
The eigenstates are multifractal, namely, the wavefunction moments
 $P_q=\int d^d r |\psi(r)|^{2q}\propto L^{-D_q(q-1)}$
present anomalous scaling with respect to the sample size $L$, where $D_q$
is a set of different exponents describing the transition \cite{wegner,mirlin}.
The second moment $P_2$ is usually referred to as the inverse participation 
 ratio (IPR).
Despite the fact that multifractal eigenstates occupy a vanishing volume,
they strongly overlap each other.
For instance, the density-density correlation function of two eigenstates
with energies $E_n$ and $E_m$ decay as a power-law,
\be
 \int d^d r \; |\psi_{n}(r)|^2|\psi_{m}(r)|^2
 ~\sim~ |E_n-E_m|^{-[1-(D_2/d)]}
\ee
(for a review see \cite{janssen,mirlin}). 

Level statistics at the AT are universal and intermediate between 
 WD and Poisson statistics.
Typical features include:
\begin{enumerate}
\item
 A scale invariant spectrum \cite{sko} such that any spectral 
 correlator utilized to describe the spectral properties does not depend
 on the system size.
 We recall that, according to the one parameter scaling theory of localization,
 the AT occurs when the disorder strength is such that the 
 conductance is size independent (the beta function has a zero).  
\item Level repulsion such that the nearest neighbor level spacing
 distribution, $P(s)$, (which gives the 
 probability of having two eigenvalues at a distance $s$) vanishes
 as $s \rightarrow 0$, thus adjacent levels still repel each other as in 
 metallic samples.  However $P(s) \sim e^{-As}$ for $s \gg 1$, similar to 
 the result for an insulator $P(s) = e^{-s}$ though the constant $A$ is always larger
 than the unity ($\sim 1.7$ at the 3D Anderson transition).
\item Long range correlators such as the number variance
 $\Sigma^2(\ell)=\langle (N_\ell -\langle N_\ell \rangle)^2 \rangle \sim \chi \ell$ for $\ell \gg 1$ 
 are asymptotically linear \cite{chi}, as for an insulator $\Sigma^2(\ell) \sim \ell$, but with a
 slope $\chi < 1$ ($\chi\sim 0.27\pm .02$ for a 3D Anderson model).
 Here $N_\ell$ is the number of eigenvalues in an interval of length $\ell$.
 Likewise the spectral rigidity (a smoothed version of the number variance)
 given by
$\Delta_{3}(\ell)=\frac{2}{\ell^4}\int_{0}^{\ell}(\ell^3-2\ell^2x+x^3)\Sigma^{2}(x)dx$
 is also linear with slope $\Delta_{3}(\ell) \sim (\chi/15) \ell$.
\end{enumerate}

The above discussion, with small modifications, holds independently of the details of the
 microscopic disorder distribution or whether the
 system has any additional symmetries.
Our case of interest, the QCD Dirac operator,
has an additional chiral symmetry which makes the matrix representing 
 the operator have a block structure given in equation (\ref{dilm}).
Obviously the additional chiral symmetry does not affect the properties in the bulk of the spectrum but 
 certainly may have an impact on the eigenvalues close to the origin. It turns out that in this 
 region \cite{nikolic,evangelou,par} 
the properties of the eigenstates depend strongly on the dimension of the space and the 
 disorder distribution. For instance for a 3D disordered system with chiral symmetry 
and short range (nearest neighbor) disorder \cite{nikolic,evangelou} 
 the eigenvectors are power-law localized with and exponent depending on the density of impurities.
 Typically,  in lower dimensions (1D,2D) 
 eigenvectors close to the origin are more delocalized than those at the bulk. The reasons for such 
 anomalous behavior have been traced back \cite{verges} to the absence
 of weak-localization corrections in the chiral case. 
By contrast in three dimensions, it 
 has been claimed \cite{nikolic} that 
eigenstates close to the origin are more localized than those in the bulk
 for a broad range of disorder strengths. However there is no consensus on whether this finding
 is universal of it is just a peculiarity of the model utilized in \cite{nikolic}.
 In any case 
 it is expected that, for a certain disorder strength, 
an AT close to the origin may coexist with the one above reported at the bulk.
 Indeed, this is also what we have found in 
 the ILM at finite temperature (see next section).
Clearly further investigation is still needed to provide a detailed account 
 of the AT close to the origin in disordered systems with chiral symmetry.

In summary,
at zero temperature certain features of the QCD vacuum resemble those of a disordered medium where the 
 role of impurities is played by an instanton liquid. The
 \SB{} is a consequence of the delocalized nature of the zero mode eigenvectors
of the QCD Dirac operator caused by the long range
 nature of the overlap matrix.  A natural question to ask is whether
 these analogies can be extended to the case of finite temperature. 
The aim to this paper is to answer affirmatively
 this question.
Specifically, in the context of an Instanton Liquid Model,
 we show that around the temperature that the chiral phase transition occurs
 the low lying eigenmodes of the QCD Dirac operator undergo an
 Anderson transition, characterized by multifractal eigenstates and
 critical statistics \cite{anderson}.

\section{Anderson localization in the ILM at finite temperature
\label{secResults}}

In this section we show that both the spectrum and the eigenvectors of
 the QCD Dirac operator in an ILM undergo a transition to localization
 at a certain temperature.
Both the quenched and the unquenched cases are considered here.
In the latter case we see that the transition to localization occurs
 around the same temperature as the chiral phase transition thus
 suggesting that both phenomena are deeply related.
This is the main result of this paper.
First we provide a short overview on how the numerical simulations
 were performed.

\subsection{Technical details of the numerical simulation}

The ILM  partition function for $N_f$ quark flavors with masses $m_k$
is given by
\be
Z_{\rm inst} = \int D\Omega  \; e^{-S_{\rm YM}}\;
 \prod_{k=1}^{N_f} {\det}(\slash{D} + m_k),
\label{zinst}
\ee
where the integral is over the positions, sizes and orientations of the
instantons
and $S_{\rm YM}$ is the Yang-Mills action.
The fermion determinant is evaluated in the space
of the fermionic zero modes of the instantons. 
For further discussion of this partition function we refer to \cite{SS97}.
We just mention that we use the standard phenomenological value of the
instanton density $N/V = 1\,\mathrm{fm}^{-4}$. For the sake 
of simplicity we have kept this density fixed as the temperature is
increased. We justify this approximation based on the 
 fact that even for temperatures close to the chiral phase
transition the instanton density is still sizable 
(around $0.6\,\mathrm{fm}^{-4}$ \cite{SS97}).
Indeed in previous simulations the drop in the density was ruled out as
 the physical mechanism leading to the chiral phase transition \cite{SS97}.

All units in the ILM are typically given in terms of the QCD scale parameter
$\Lambda$ which we have simply set to 200 MeV.
This choice is sufficient for our purposes since we are not concerned with
making quantitative predictions about the position of the QCD chiral
phase transition, but instead are interested in more general features of
the transition that we don't expect to be very sensitive to a fine tuning
of the parameters of the ILM.
We therefore use the
ILM as a qualitative model for QCD at
temperatures around the chiral phase transition.
We also stress that it provides a reasonable
 description of \SB{} and many hadronic correlation functions both at zero
 and at finite temperature.
Furthermore the ILM has modest computational requirements and allows
 us to go to fairly large volumes and get good statistics. 

In this paper we present results for both the quenched $N_f = 0$ and
the unquenched $N_f=2,3$ cases ($N_f=2$ with $m_u=m_d=0$ and
at $N_f=3$ with $m_u=m_d=20$ MeV and $m_s=140$ MeV).
The partition function, equation (\ref{zinst}), is evaluated using a standard
Metropolis algorithm.
In the quenched case we performed between 2000 to 10000 measurement
sweeps for each set of parameters after allowing around 1000 sweeps
for thermalization.
Results are presented for ensembles of up to $N=5000$ instanton and
anti-instantons.
In the unquenched case we were only able to
investigate ensembles of up to $N=700$ instantons and anti-instantons.
We also discarded the first 1000 sweeps in each simulation and did between
1000 to 5000 sweeps per ensemble.
Thermodynamic properties of the model have already been addressed in
\cite{SS97} and will not be discussed here.
We will focus mainly on observables 
such as level statistics and eigenvector scaling properties of the
 QCD Dirac operator which are especially well suited
 for studying Anderson localization effects \cite{mirlin}.
All the spectral correlators are calculated from the unfolded spectrum.
This procedure scales the eigenvalues so that the spectral 
density on a spectral window comprising
several level spacings is unity but 
it does not remove the small fluctuations about the mean density that
provide fundamental information about the system.

Finally we recall that in the ILM the effective disorder parameter is the
temperature since we have fixed the density of instantons. One 
of the most challenging tasks of the numerical calculation is to 
accurately locate the temperature 
at which the AT occurs. In all cases in the 
paper this temperature has been estimated by 
the finite size scaling method introduced in \cite{sko} 
in the context of disordered systems.
In essence this method consists of computing a spectral correlator, such as the
level spacing distribution $P(s)$ or the number variance $\Sigma^2(\ell)$, 
for different sizes and then finding the temperature at which it becomes size
independent.

\begin{figure}[t]
 \hfill
 \begin{minipage}[t]{.48\textwidth}
  \begin{center}
   \includegraphics[width=\columnwidth,clip,angle=0]{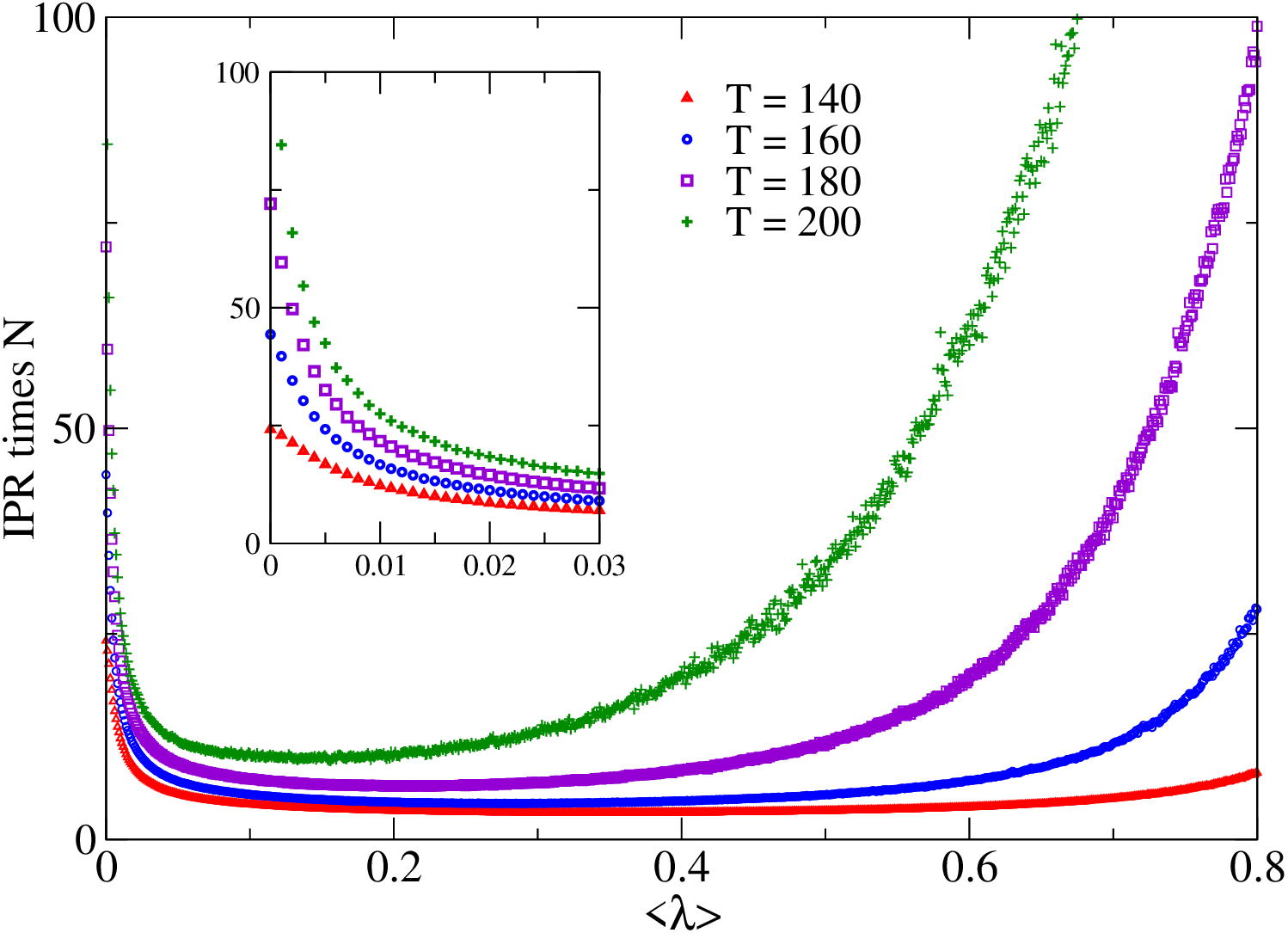}
%   \vspace{3mm}
   \caption{Inverse participation ratio (IPR) times $N$ versus the average
    eigenvalue
    for the quenched ILM with $N=2000$ instantons at different temperatures. 
    Low lying eigenmodes are more localized than those in the bulk for all 
    tested temperatures.}
   \label{nf0ipr}
  \end{center}
 \end{minipage}
 \hfill
 \begin{minipage}[t]{.48\textwidth}
  \begin{center}
   \includegraphics[width=\columnwidth,clip,angle=0]{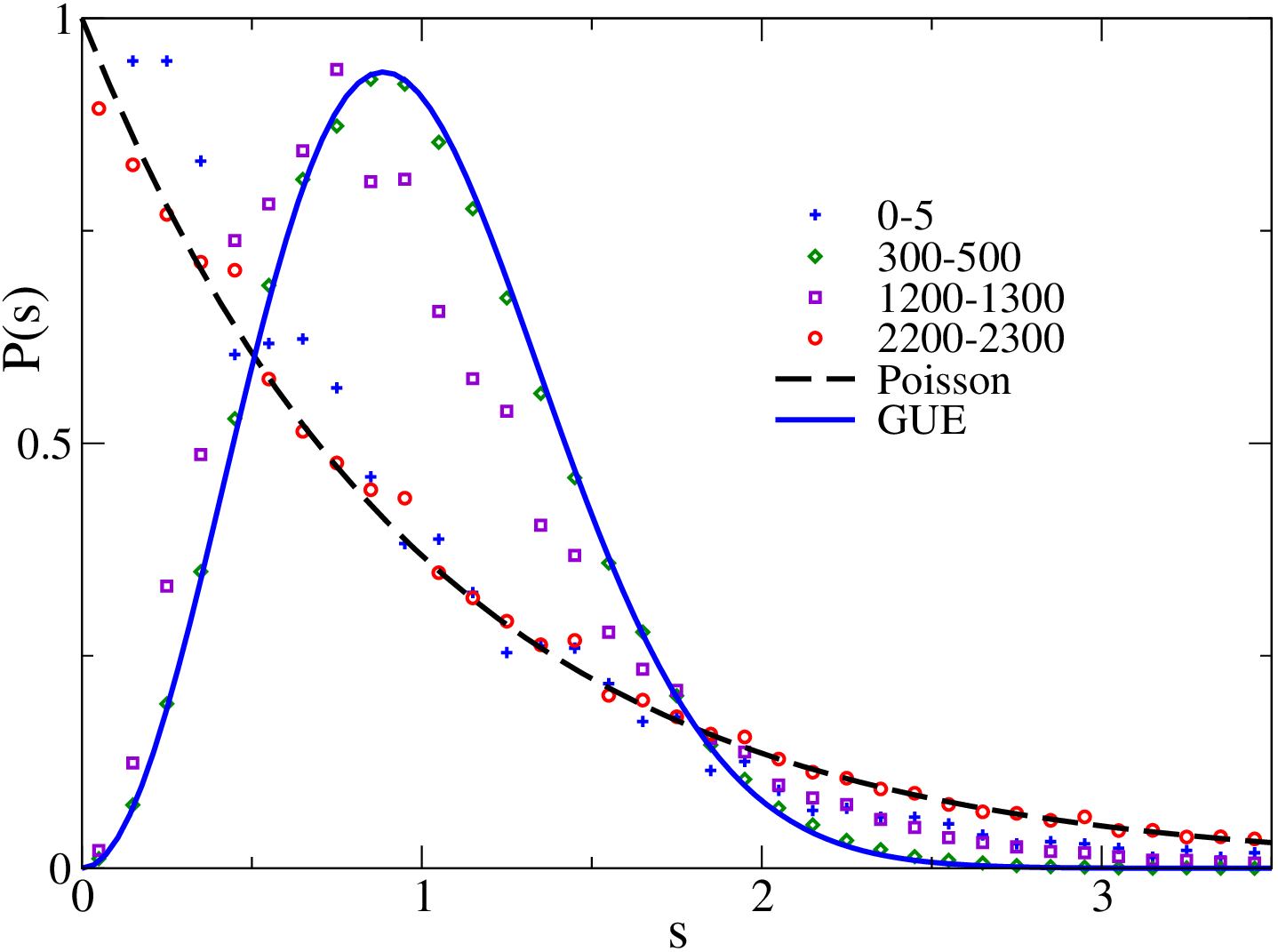}
%   \vspace{3mm}
   \caption{Level spacing distribution $P(s)$ for the quenched ILM with $N = 5000$
      instantons at
    $T=200$ MeV for different spectral windows.
    The legend indicates the eigenvalue numbers with 0 being the
    smallest one.
    The spectral correlations of the lowest and highest modes are close to
       Poisson statistics. By contrast the spectral 
      correlations in the central 
     part of the spectrum are well described by RMT (GUE result).}
   \label{nf0ps}
  \end{center}
 \end{minipage}
 \hfill
\end{figure}

\subsection{Anderson localization in the ILM: The quenched case}

In this section we present results for the quenched ILM
at nonzero temperature.
Without dynamical fermions there is not technically a chiral phase
transition since there is no chiral symmetry.
However one can in principle study
the ``quenched'' quark condensate in a purely gluonic background 
 through the Banks-Casher relation \cite{bank}.
Lattice simulations with staggered fermions \cite{lattrans}
 have also found that the vanishing of the ``quenched'' quark condensate
 occurs at the same temperature as the deconfinement transition does.
Since the ILM does not have a confinement-deconfinement transition \cite{SS97}
 we can only speak about the chiral phase transition in the ILM.
However, in the quenched ILM, the spectral density seems to diverge close
 to the origin
 (similarly to recent lattice results with overlap fermions \cite{kiskis})
 thus suggesting a likewise divergent ``quenched'' quark condensate.
In view of these facts the transition to localization (see below) 
 we have observed in the quenched QCD Dirac operator at finite temperature
 both close to origin and in the bulk of the spectrum 
 can't be linked directly to either of these phenomena. 
We still consider the quenched ILM results to be of interest 
 as we have found a very clear example of a mobility edge
 with features strikingly similar to those of a 3D disordered system
 at the AT. This finding reinforces the idea \cite{aj}
 that certain aspects of the QCD vacuum resemble those of a disordered system. 

\begin{figure}[t]
 \hfill
 \begin{minipage}[t]{.48\textwidth}
  \begin{center}
   \includegraphics[width=\columnwidth,clip,angle=0]{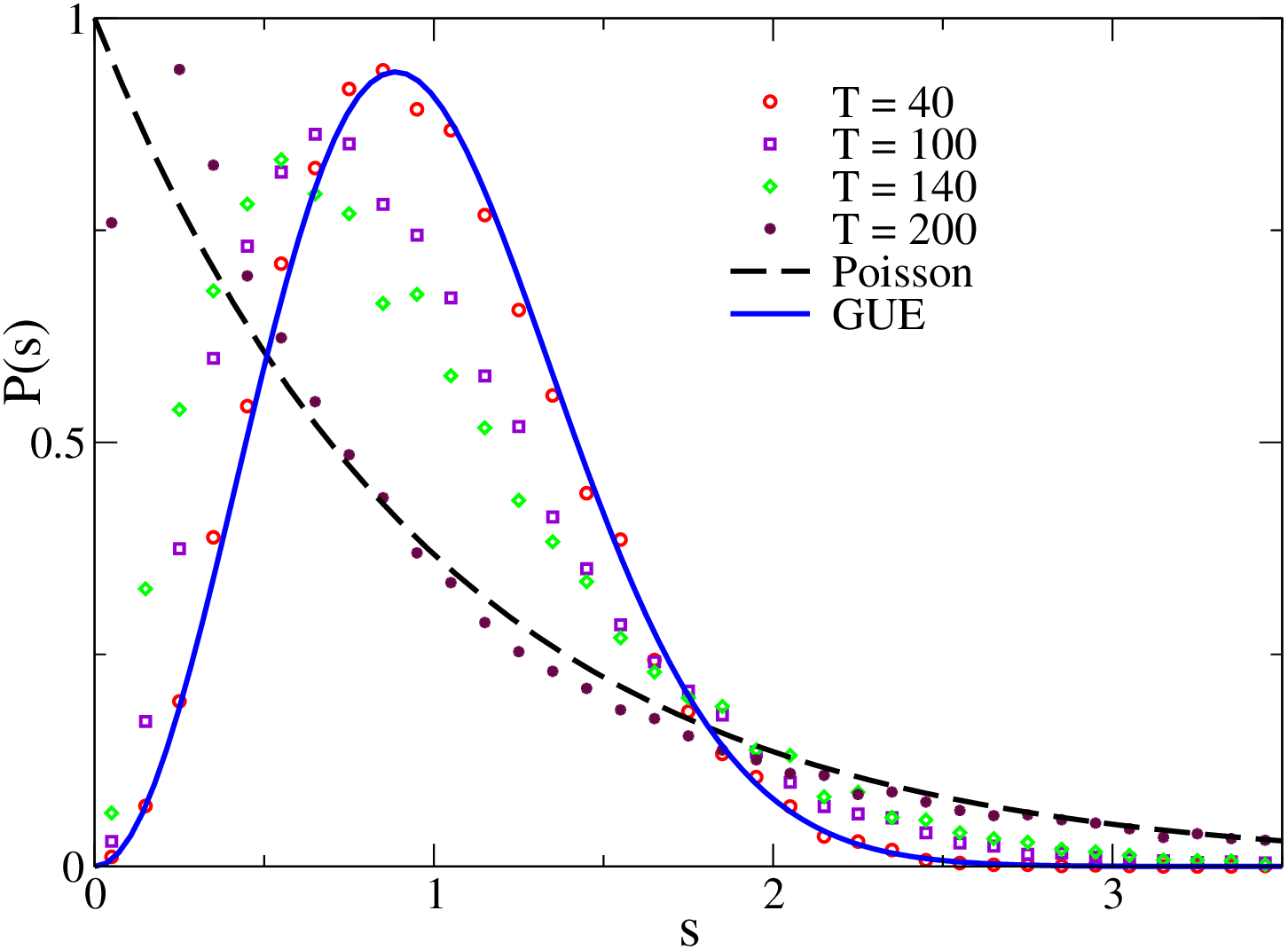}
%   \vspace{3mm}
   \caption{Level spacing distribution $P(s)$ close to the origin 
     of the spectrum for the
    quenched ILM with $N = 2000$ instantons for different temperatures.
    Agreement with the prediction of RMT (GUE result) is observed for relatively
    low temperatures $T < 100$ MeV.
    A transition between RMT (GUE) and Poisson statistics is observed in the
    range $T\sim 100-140$ MeV.}
   \label{nf0pso}
  \end{center}
 \end{minipage}
 \hfill
 \begin{minipage}[t]{.48\textwidth}
  \begin{center}
   \includegraphics[width=\columnwidth,clip,angle=0]{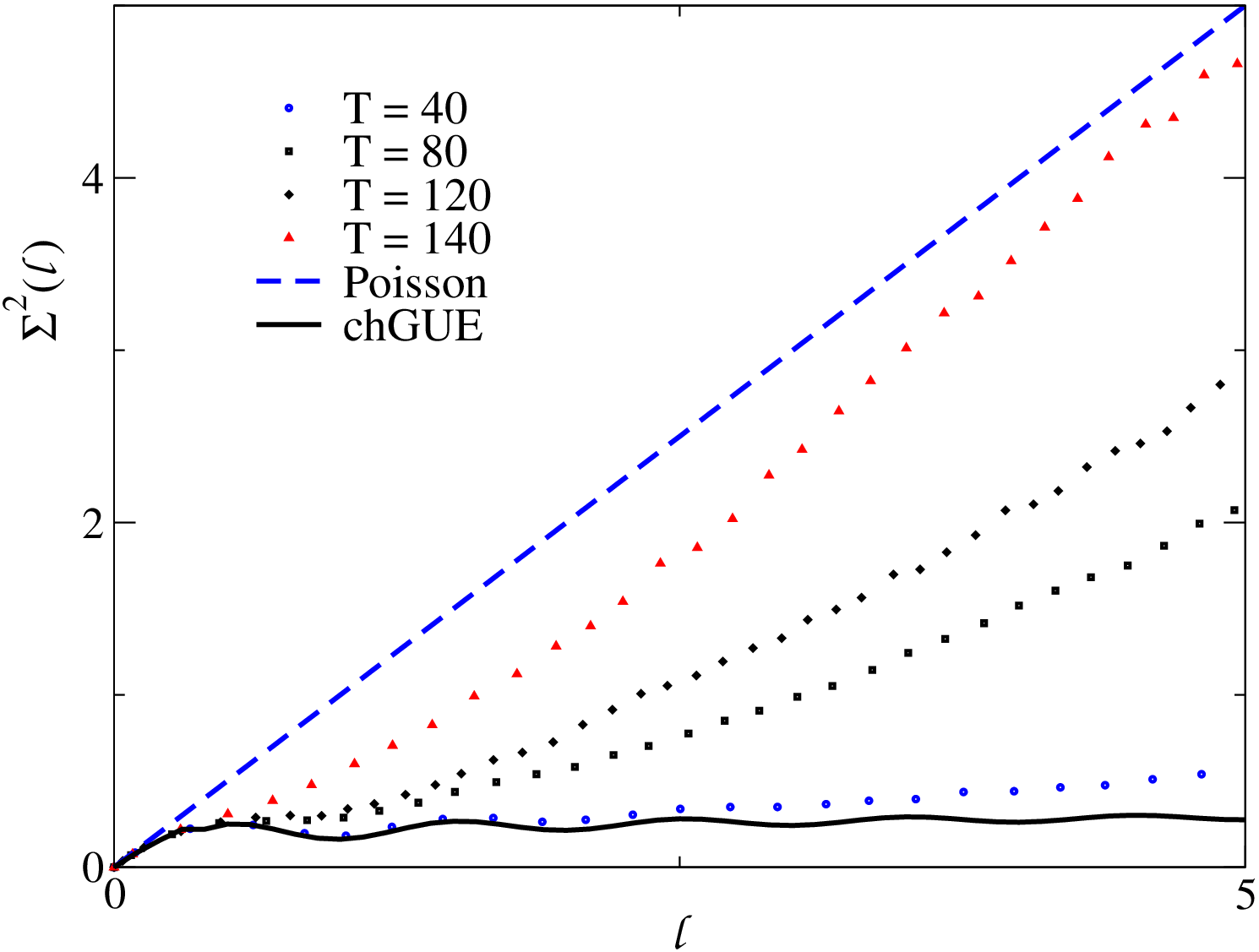}
%   \vspace{3mm}
   \caption{Number variance $\Sigma^2(\ell)$ at the origin for the
    quenched ILM with $N =1600$ instantons for different temperatures.
    Agreement with the RMT (chiral GUE) result is observed for
    relatively low temperatures $T < 80$ MeV.
    The temperature range of the transition to Poisson agrees with that
    obtained from the level spacing distribution.}
   \label{nf0nvo}
  \end{center}
 \end{minipage}
 \hfill
\end{figure}

We start by summarizing our main findings:
\begin{enumerate}
\item
The eigenvectors of the QCD Dirac operator at finite temperature 
with the lowest eigenvalues
 (the origin) are much more localized than the ones in the middle
 of the spectrum (the bulk).
Signs of stronger localization close to the origin were observed not only 
in the eigenvector moments
 (figure \ref{nf0ipr}) but also in the level statistics (figure \ref{nf0ps}).
 This difference between the origin and bulk has also been reported in 3D
 disordered systems with chiral symmetry \cite{nikolic}.
\item
 In the bulk we find a mobility edge separating localized from
 extended states which moves toward the high end of the spectrum as
 the temperature is decreased.
 The level statistics of eigenstates around the mobility edge 
 are well described by critical statistics
 (see figures \ref{nf0psb} and \ref{nf0d3b})
 typical of a 3D disordered conductor at the AT.
Outside the critical region the level statistics are
 similar to those of a metal (Wigner-Dyson statistics)  
 in the region below the mobility edge. By contrast,
 in the region above the mobility edge, it resembles those
 of an insulator (Poisson statistics).
\item
 Close to the origin
 we observe a transition to localization for temperatures in the range
 $T\sim 100 - 140$ MeV (see figure \ref{nf0pso}).
\end{enumerate}

\subsubsection{The bulk}

\begin{figure}[t]
 \hfill
 \begin{minipage}[t]{.48\textwidth}
  \begin{center}
   \includegraphics[width=\columnwidth,angle=0]{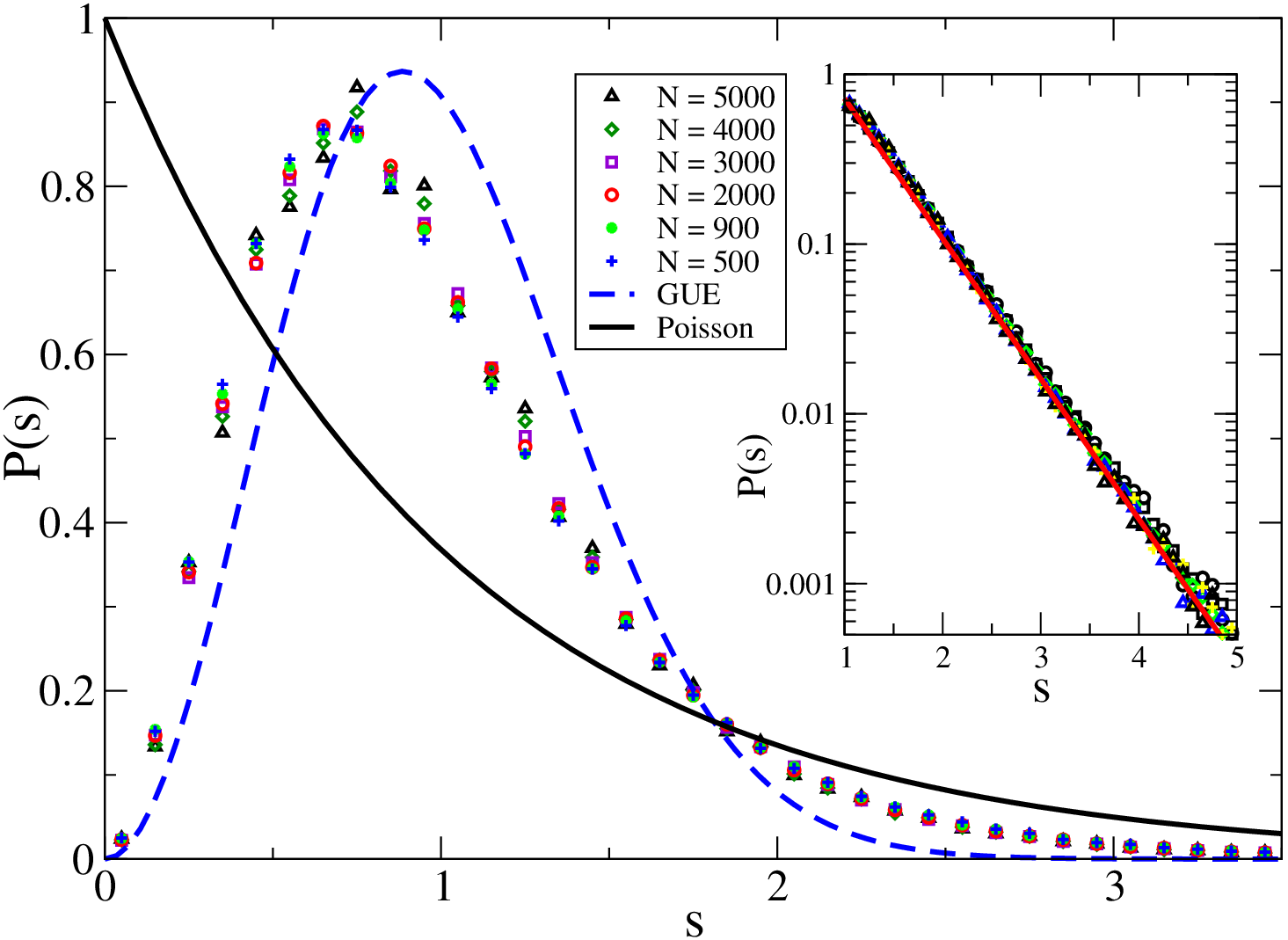}
%    \vspace{3mm}
   \caption{Level spacing distribution $P(s)$ in the bulk of the spectrum 
        for the quenched
            ILM at $T=200$ MeV and different numbers of instantons.
            The inset shows the tail of $P(s)$.
            The solid line corresponds to the best fit $P(s) \sim e^{-1.86s}$.}
   \label{nf0psb}
  \end{center}
 \end{minipage}
 \hfill
 \begin{minipage}[t]{.48\textwidth}
  \begin{center}
   \includegraphics[width=\columnwidth,angle=0]{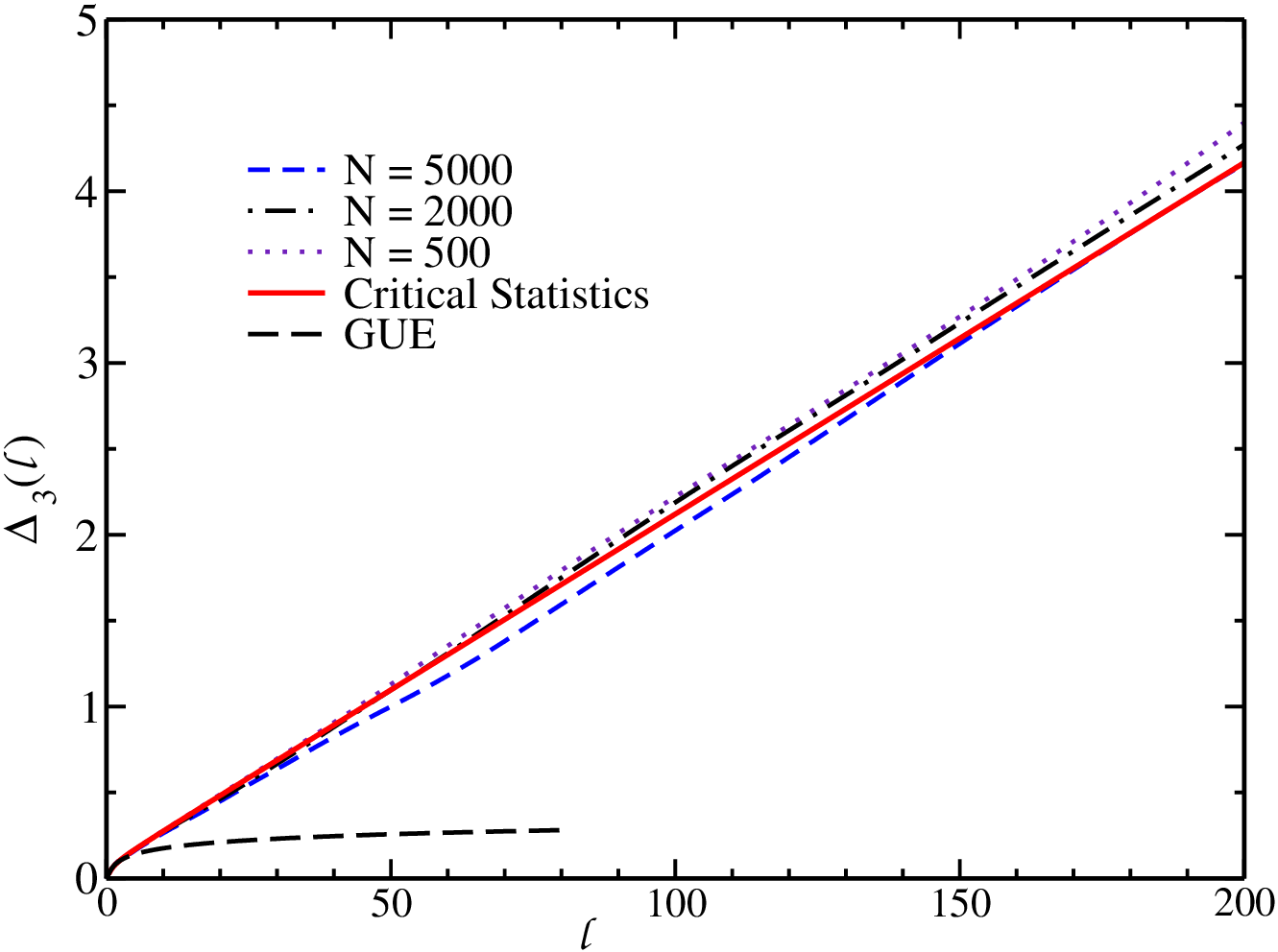}
%   \vspace{3mm}
   \caption{Spectral rigidity $\Delta_3(\ell)$ in the bulk of the spectrum 
        for the quenched ILM at $T=200$ MeV
            and different numbers of instantons.
            The result is almost size independent and agrees well with
            the prediction of critical statistics (solid line).}
   \label{nf0d3b}
  \end{center}
 \end{minipage}
 \hfill
\end{figure}

Using the finite scaling method introduced in \cite{sko} we found a
 mobility edge in the bulk of the spectrum
in the range $T \sim$ 150 -- 250 MeV.
As the temperature
decreases its location moves to the end of the
spectrum. For $T < 170$ MeV the results are less reliable since the
mobility edge is located almost at the end of the spectrum where
truncation effects due to the finite size of the ILM are larger.
Following the literature in
disordered systems we have investigated the temperature $T \sim 200$ MeV
such that the mobility edge is located around the center of the
spectrum.
We initiate our analysis by looking at the level statistics.
We recall the spectral
correlators were computed from the unfolded spectrum
 in the region around the mobility edge.
The main findings are summarized as follows:
\begin{enumerate}
\item
 The spectrum is scale invariant to high degree. As observed in
 figure \ref{nf0psb},
 the level spacing distribution $P(s)$ does not depend much on the system size
 for volumes ranging from $N=500$ to $N=5000$. The inset plot shows
 that even the tail of $P(s)$ which contains information about
 correlations at larger scales is also scale invariant. The study
 of long range correlators further confirms
 this behavior.  Likewise, in figure \ref{nf0d3b}, it is shown that the
 spectral rigidity varies little with the system size in a window of more
 than $200$ eigenvalues around the mobility edge. 
\item Level repulsion,  $P(s) \rightarrow 0$ as $s \rightarrow 0$, typical 
 of a metal is still
 present (figure \ref{nf0psb}).
 However $P(s)\sim e^{-1.86s}$ has an exponential tail as for
an insulator. Both features also appear in the case of a 3D disordered system 
 at the AT.
\item The spectral rigidity (figure \ref{nf0d3b}) is asymptotically
 linear with a slope $\chi/15 \sim
 (0.29 \pm 0.02)/15$ in fair agreement with the result of a 3D disordered
 system at the AT of $\chi \sim 0.27/15$ \cite{mirlin}. 
\item The level statistics of the ILM in the critical region are accurately 
 described by generalized random matrix models \cite{chen,ant3} capable of
 reproducing critical statistics (see figure \ref{nf0d3b}).  
The explicit analytical result for the
 chiral case can be found in \cite{kazu} 
for a random banded model with a power-law decay 
and in \cite{ant3,tsv} in the context of the Calogero-Sutherland model
 \cite{calo} at finite temperature.  We recall that in these models
 there exits a free parameter that must be fixed by, for instance,
 matching the slope of the spectral rigidity $\chi$. 
In our case, as shown in figure \ref{nf0d3b}, 
the exact form of the spectral rigidity in the bulk
 follows closely the prediction of critical statistics, equation (31) in
 \cite{ant3} with $h$, a free parameter, set to a value of $0.62$.
\item The level statistics outside the AT window are
 given by Wigner-Dyson statistics (typical of a metal) in the region below the
 mobility edge. By contrast it is close to Poisson statistics typical of an
 insulator in the region above the mobility edge.
 This is corroborated in figure \ref{nf0ps} where we observe
 a transition from Poisson to Wigner-Dyson statistics in $P(s)$ at
 $T=200$ depending the region of the spectrum where $P(s)$ is computed.
\end{enumerate}
The above analysis provides 
compelling evidence that for a certain range of temperatures 
the level statistics in the bulk of the QCD Dirac operator in the
 ILM are consistent with those of a 3D disordered system at the AT.

We now turn to eigenvector properties.
As mentioned previously, a signature of the AT is the multifractality of the
eigenstates. Multifractality is usually detected 
 by the anomalous scaling of the moments of the eigenfunction $P_q$
(defined in section \ref{secpq}) with the system size $L \propto N^{1/3}$.
We have computed the fractal dimension $D_2$ by
fitting $P_2$ for different system sizes $N$.
A fit to the form $P_2 = a N^b$ 
yields $b \sim 0.50(5)$ and consequently $D_2 \sim 1.5(1)$, consistent with
 the $D_2 \sim 1.4(2)$ found for a disordered system at the 3D AT.
In order not to consider eigenstates
outside the critical region we have taken only $10\%$ of the
eigenvectors around the center of the spectrum.

To recapitulate, the numerical analysis shows with great accuracy that at
$T=200$ MeV the QCD Dirac operator in a background field given by the
quenched ILM
undergoes and metal-insulator transition in the central part of the spectrum.
The eigenstates around the transition region are fractal and the level
statistics are well described by critical statistics, namely, they have
all the properties expected in an AT: 
scale invariance, level repulsion and sub-Poisson spectral rigidity.
Furthermore the numerical value of parameters such as the fractal
dimension $D_2$ and the slope of the number variance $\chi$ are very 
close to the results for a standard 3D disordered system at the AT. 

\subsubsection{The origin}

This region is of special interest 
since the smallest eigenmodes are responsible for \SB. However 
as mentioned earlier, in the quenched case there is not, in a strict sense,
 chiral symmetry so the localization properties in this region 
cannot be related neither to \SB{} nor its restoration at finite temperature.
Despite these limitations we have also investigated
both the spectrum and the eigenmodes in this region. Our main motivation 
 is to test whether the similarities between
 the QCD vacuum (as given by the ILM at a certain temperature) and a 3D
 disordered system at the AT  
 can also be extended to region close to the origin thus reinforcing the
 idea put forward in \cite{aj} that
 both models belong to the same universality class.

As in the bulk, the level statistics are temperature dependent.
For low temperatures $T < 80$ MeV (see figures \ref{nf0pso} and \ref{nf0nvo}), 
the eigenmodes are delocalized and the
level statistics are very close to the RMT prediction as given by the GUE. 
In the region $T \sim 100-140$ MeV we observe a transition to 
 localization.
By contrast at higher temperatures the spectrum is very nearly Poisson
 like as for an insulator.

We remark that the numerical results close to the origin
 are less conclusive than those in the bulk. 
On the one hand the analysis is complicated by the accumulation of very small
eigenvalues present in the quenched ILM. On the other hand the 
 statistics are worse since only the smallest eigenvalues are
 affected by the chiral symmetry of the model. 
Despite these technical problems it seems clear that the QCD Dirac operator
 undergoes a transition to localization for $T \sim 120$ MeV.
 As shown in figure \ref{nf0ipr},
eigenvectors for the smallest eigenvalues, due to the additional chiral
symmetry, are more localized than those in the bulk. This has already been
observed in lattice simulations \cite{damgard,lattice} and 3D
disordered systems with chiral symmetry \cite{nikolic}. We do not have a clear
understanding of the reason for such behavior. Indeed 1D and 2D chiral
disordered models usually show the opposite behavior, namely, states in the
bulk are localized for any amount of disorder and only the states close to
the origin (the ones affected by the chiral symmetry) are truly
delocalized.

\begin{figure}[t]
 \hfill
 \begin{minipage}[t]{.48\textwidth}
  \begin{center}
   \includegraphics[width=0.9\columnwidth,clip,angle=0]{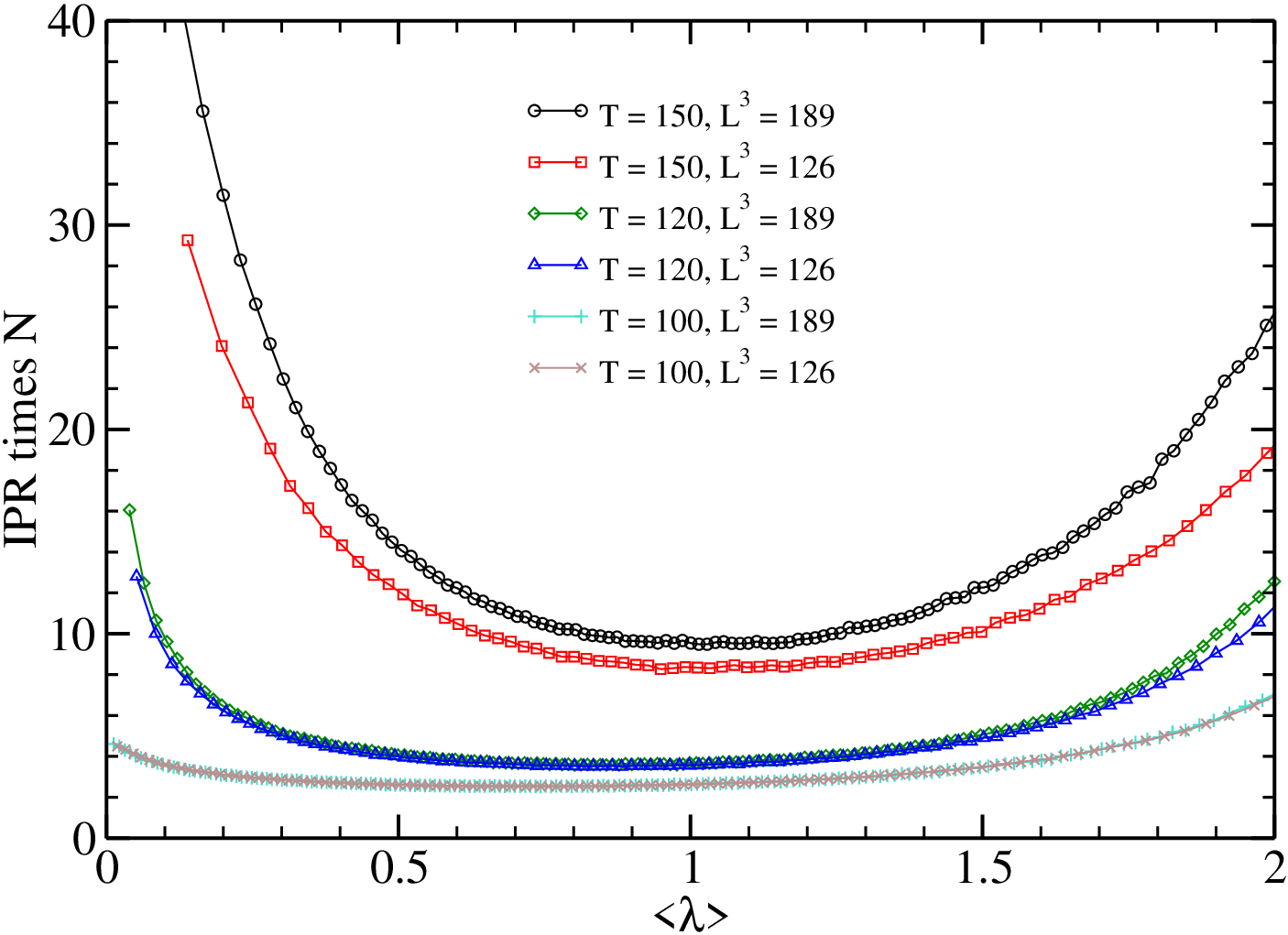}
%   \vspace{3mm}
   \caption{IPR times the total volume ($V=N$ in units of fm) versus the
            average eigenvalue
            for the ILM with 2 massless quarks at different temperatures
            and spatial volumes ($L^3$, in units of fm$^3$).}
   \label{iprnf2m0}
  \end{center}
 \end{minipage}
 \hfill
 \begin{minipage}[t]{.48\textwidth}
  \begin{center}
   \includegraphics[width=\columnwidth,clip,angle=0]{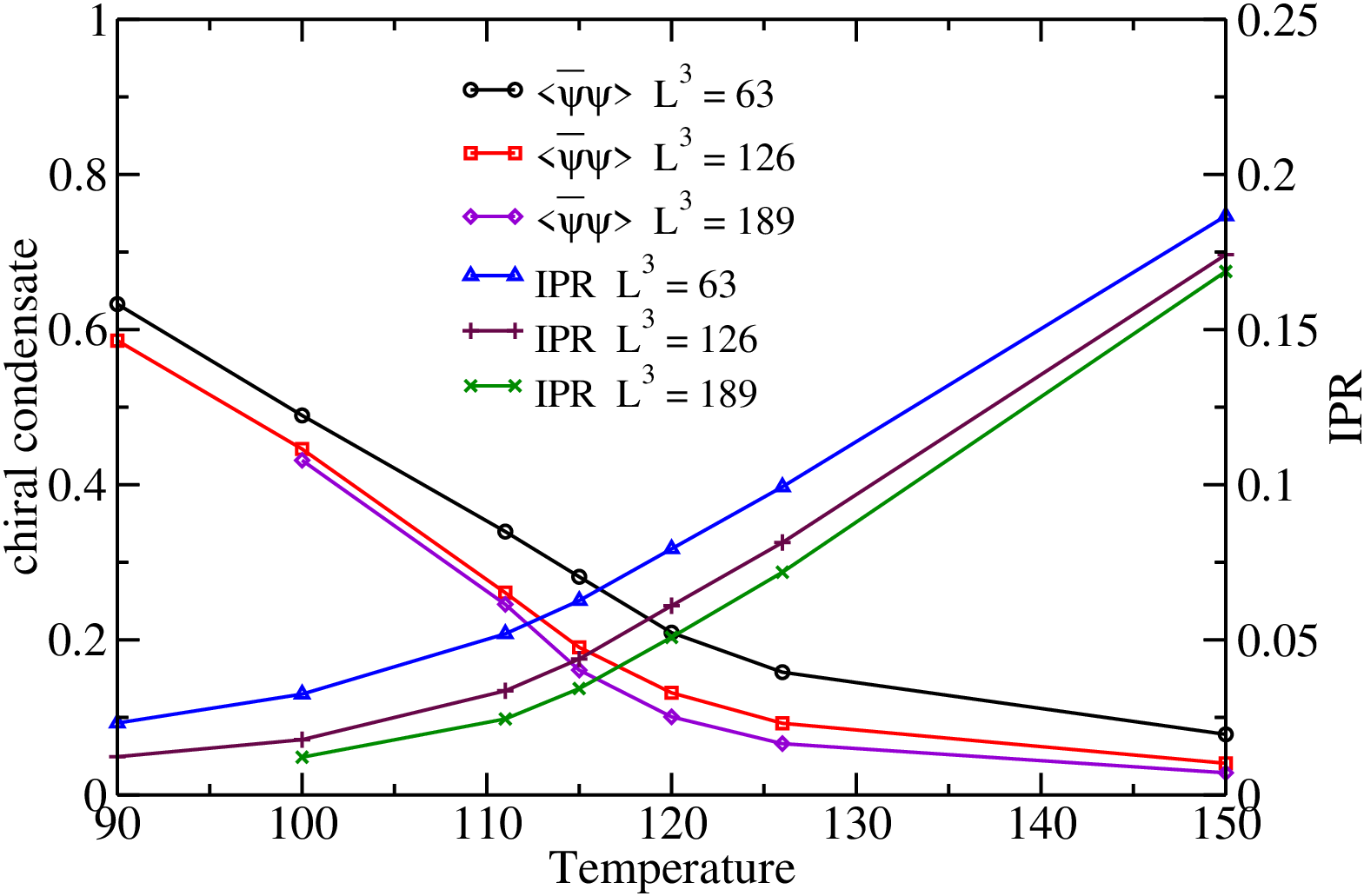}
%   \vspace{3mm}
   \caption{Chiral condensate $\langle {\bar \psi}\psi \rangle$ and the IPR 
       of the lowest eigenmode versus
            temperature for the ILM with 2 massless quarks at different
            spatial volumes ($L^3$, in units of fm$^3$). 
             Remarkably the drop in the condensate occurs in the same range of temperatures as the 
        transition to localization of the lowest eigenmodes of the QCD Dirac operator.}
   \label{cciprnf2m0}
  \end{center}
 \end{minipage}
 \hfill
\end{figure}

\subsection{Anderson transition and chiral phase transition:
            The unquenched case}

In this section we investigate the eigenvalue and eigenvector
statistics of the QCD Dirac operator with two massless quark flavors
and for three flavors with $m_u=m_d=20$ MeV and $m_s=140$ MeV. 
In the massless case a true chiral phase transition is supposed to
occur at a certain (chiral restoration) temperature.
Our main aim is to investigate whether the chiral phase transition and
the localization transition in the QCD Dirac operator occur at the
same temperature.
Below we show that this is indeed the case thus suggesting that the
phenomenon of Anderson localization may be the fundamental mechanism
driving the chiral phase transition.

We mention that in the massive case there is not a chiral phase transition
but instead a crossover as is believed to occur in full QCD.
The quark masses have been chosen larger than the experimental values
in order to observe this crossover in the ILM.
Additionally such large mass values are closer to the ones typically
used in current lattice calculations.

The behavior of the ILM at finite temperature for different numbers of
flavors has been studied previously \cite{SS96}.
The case of just one flavor $N_f=1$ is special since
the chiral symmetry is explicitly
broken by the $U(1)$ anomaly, a condensate is formed but is not
related to the spontaneous breaking of any symmetry. This condensate
decreases smoothly with the temperature and there is no sign of any
phase transition \cite{SS96}. By contrast for $N_f>5$
the chiral symmetry is no longer broken (in the chiral limit) at
zero temperature. This is due to the fact that the fermionic
determinant in the partition function suppresses configurations with
low-lying eigenvalues. The rate of suppression increases with the
number of flavors as $\lambda^{N_f}$. It is believed that for $N_f \sim 5$
the spectral density of the Dirac operator develops a gap close to the
origin and, by the Banks-Casher relation, the chiral symmetry is not
broken even at zero temperature.

\begin{figure}[t]
 \centering
 \begin{minipage}[c]{0.48\columnwidth}
  \centering
  \includegraphics[width=0.94\columnwidth,clip]{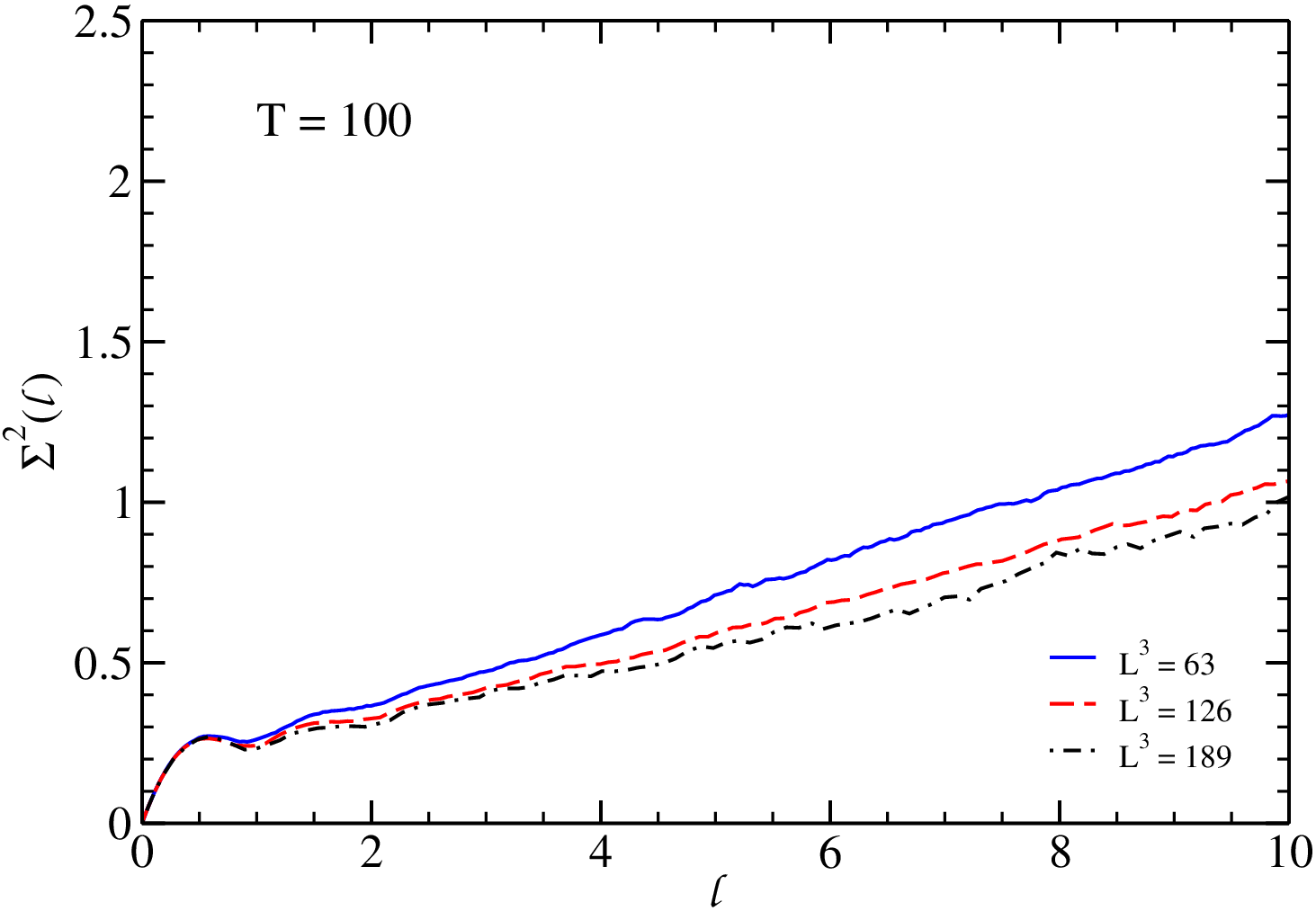}
  \includegraphics[width=0.94\columnwidth,clip]{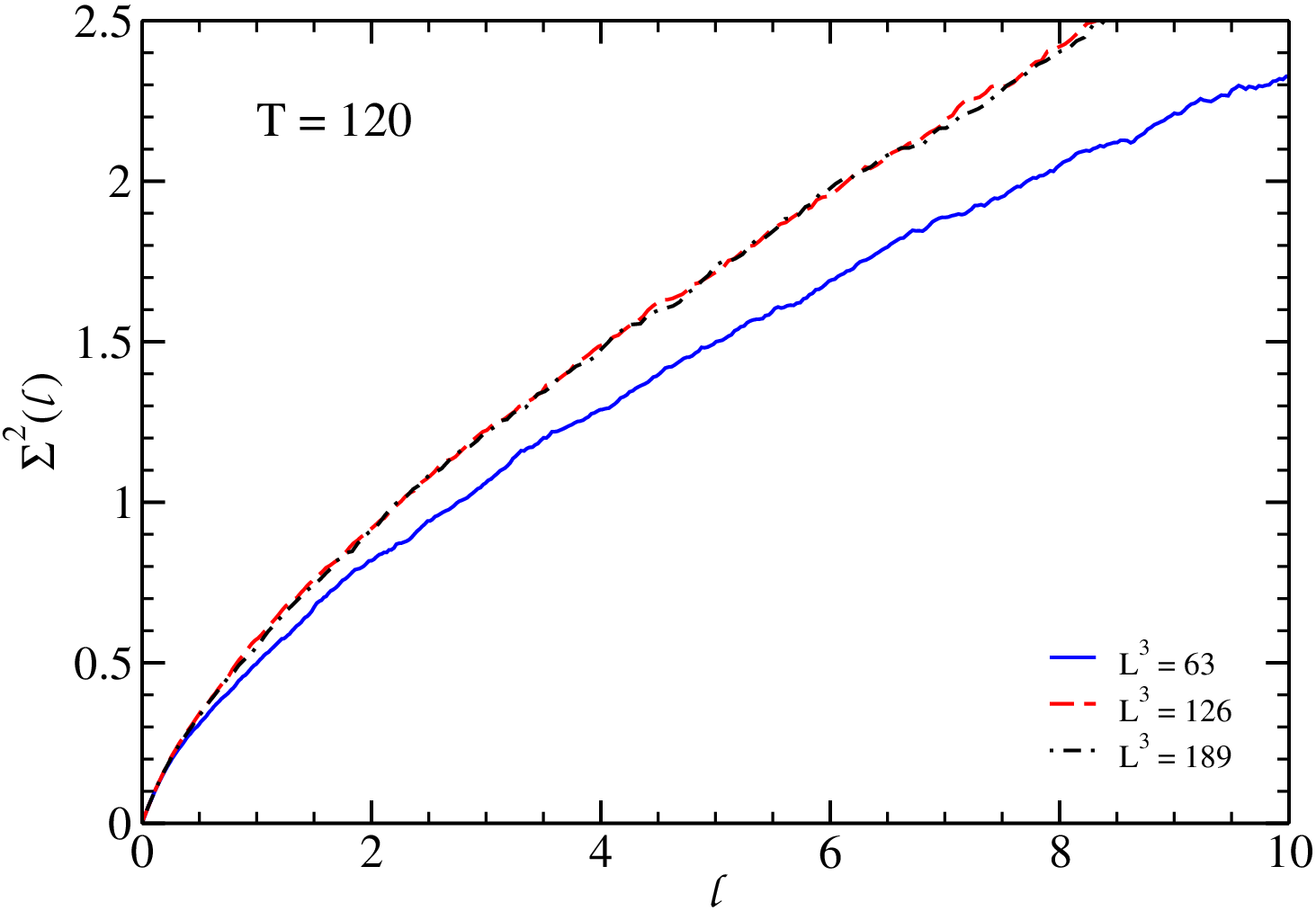}
 \end{minipage}
 \hfill
 \begin{minipage}[c]{0.48\columnwidth}
  \centering
  \includegraphics[width=0.94\columnwidth,clip]{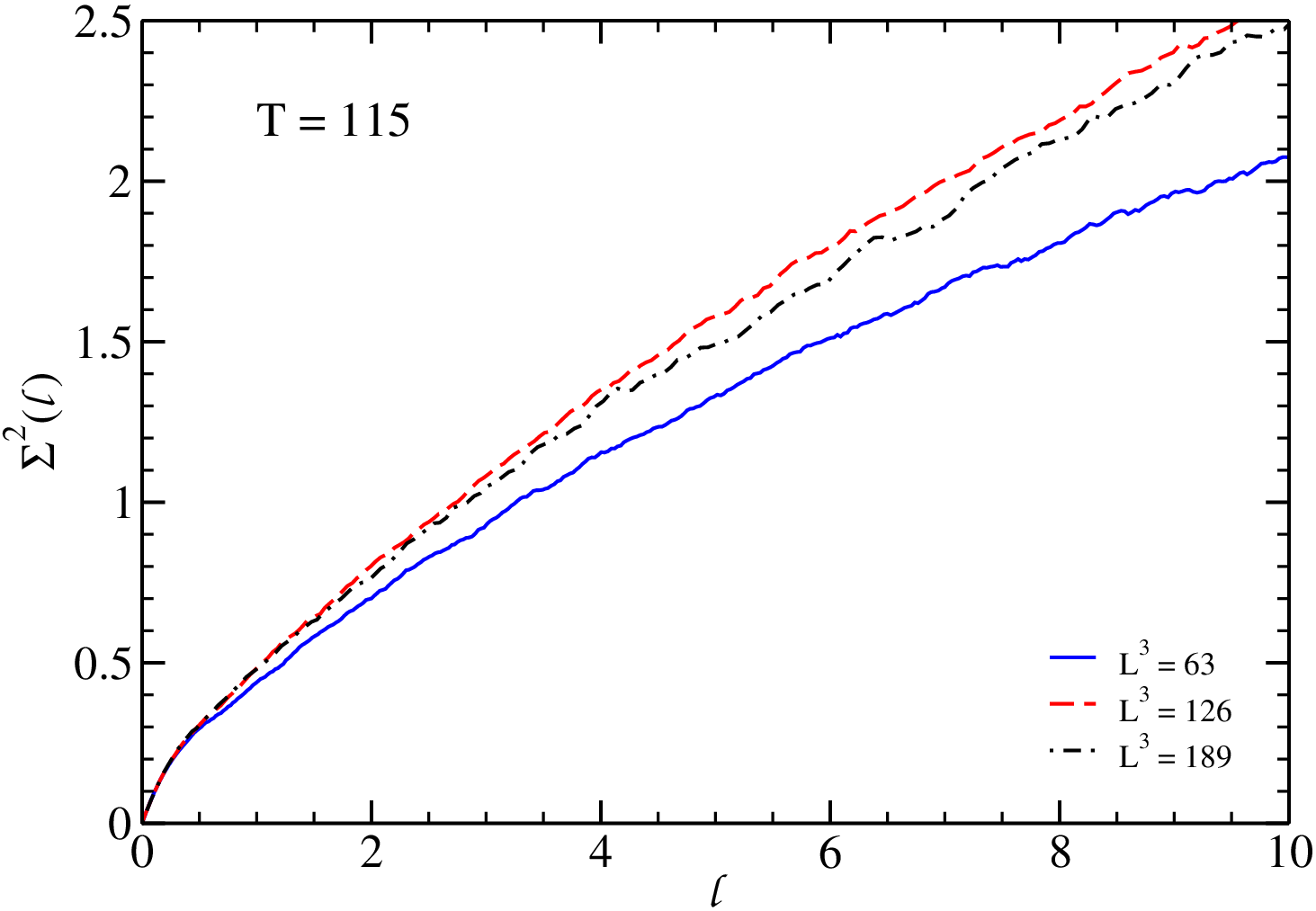}
  \includegraphics[width=0.94\columnwidth,clip]{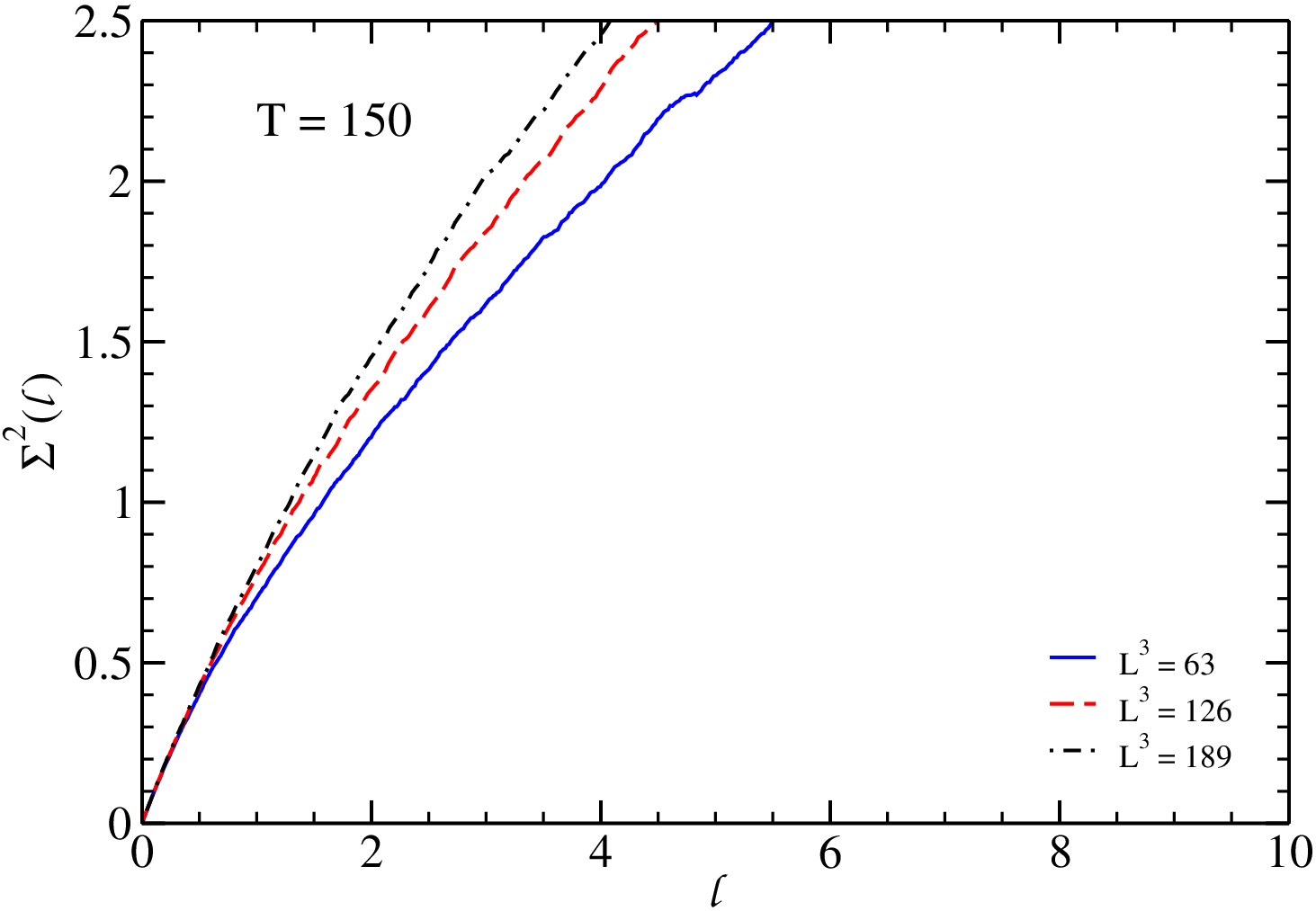}
 \end{minipage}
% \vspace{3mm}
 \caption{Number variance $\Sigma^2(\ell)$ at the origin for the ILM with
          two massless quark flavors at different temperatures and spatial
          volumes ($L^3$, in units of fm$^3$).
          The scaling with volume is very small for $T \sim 120$ MeV which
          indicates that the critical temperature associated with the AT is
          very close.}
 \label{nvonf2m0}
\end{figure}

We start our analysis with the eigenmode properties of the QCD Dirac operator
with two massless flavors. 
In figure \ref{iprnf2m0} we show the IPR times the total volume $V$ (or number of instantons $N$) versus
the average eigenvalue for different temperatures and spatial volumes.
Completely extended states will have a constant IPR$\times V$ 
independently of the system size,
while for completely localized states IPR$\times V$ 
will be proportional to the volume.
As in the quenched case
eigenstates close to the origin are more localized than those in the
bulk but the difference between the two regions is  smaller than in
the quenched case. This is not surprising since the fermionic
determinant suppress low-lying eigenvalues thus weakening the
chiral symmetry and consequently the difference between the origin and
bulk.
At the lowest temperature ($T=100$ MeV) we clearly see that the scaled IPR
is independent of the volume and consequently the eigenmodes are delocalized.
The opposite occurs for $T=150$ MeV where scaling is observed
in all eigenmodes, especially the ones close to the origin, indicating that 
the eigenmodes are localized.
An intermediate situation 
 happens at $T \sim 120$ MeV where 
we see some volume dependence near the origin and at the
high end of the spectrum while the bulk states remain fairly constant.
This indicates that a mobility edge separating localized 
 from delocalized states is forming at the origin. A natural 
question to ask at this point is whether this transition
 to localization occurs in the same range of temperatures
 as the chiral phase transition.

\begin{figure}[t]
 \hfill
 \begin{minipage}[t]{.48\textwidth}
  \begin{center}
   \includegraphics[width=\columnwidth,clip,angle=0]{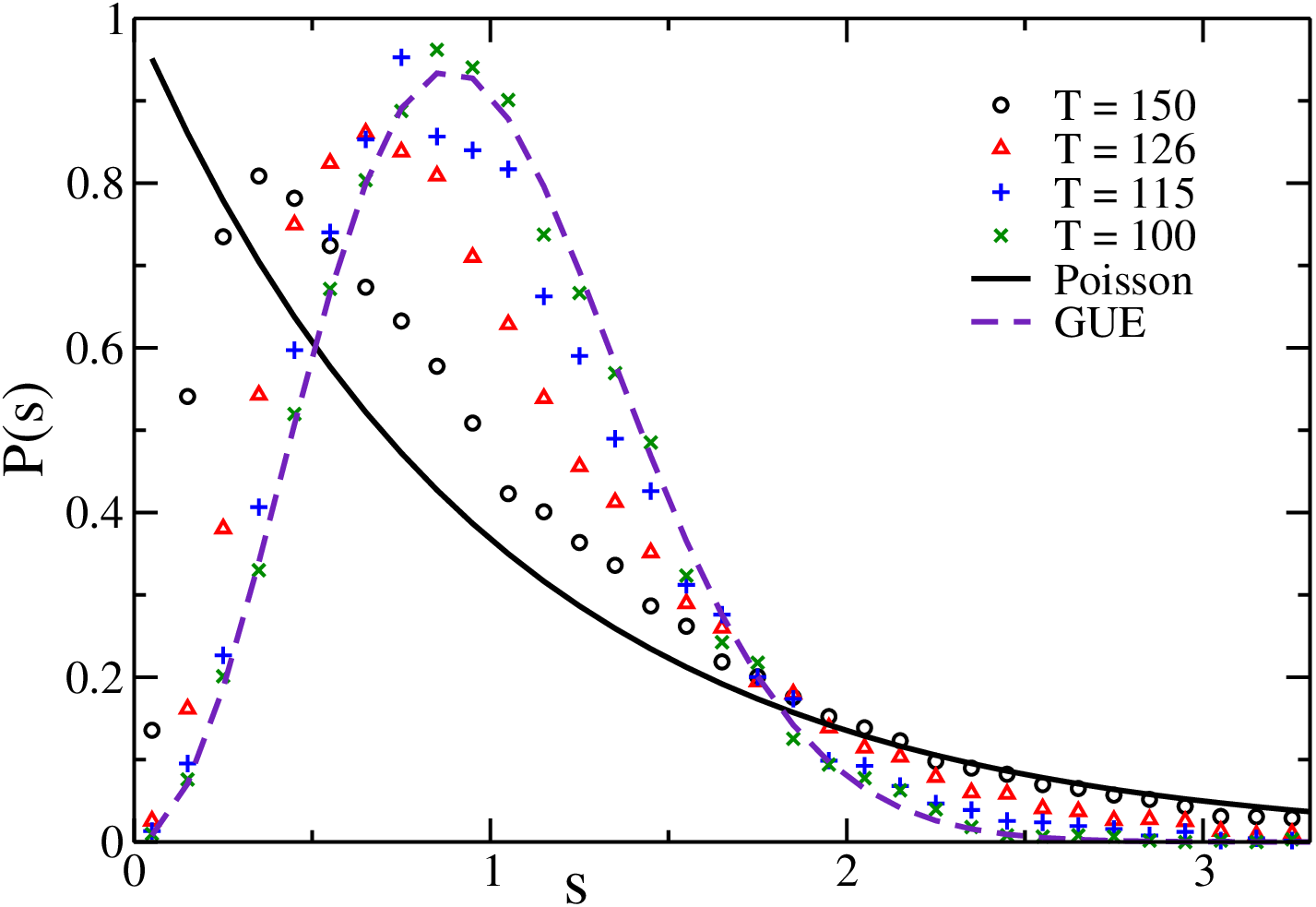}
%   \vspace{3mm}
   \caption{Level spacing distribution $P(s)$ at the origin for the ILM with 2
            massless quarks with spatial volume $L^3=189\,\mathrm{fm}^3$ for
            different temperatures. A transition from the RMT prediction (GUE) typical of a metal
 to the Poisson result typical of an insulator is observed as the temperature is increased.}
   \label{psnf2m0}
  \end{center}
 \end{minipage}
 \hfill
 \begin{minipage}[t]{.48\textwidth}
  \begin{center}
   \includegraphics[width=\columnwidth,clip,angle=0]{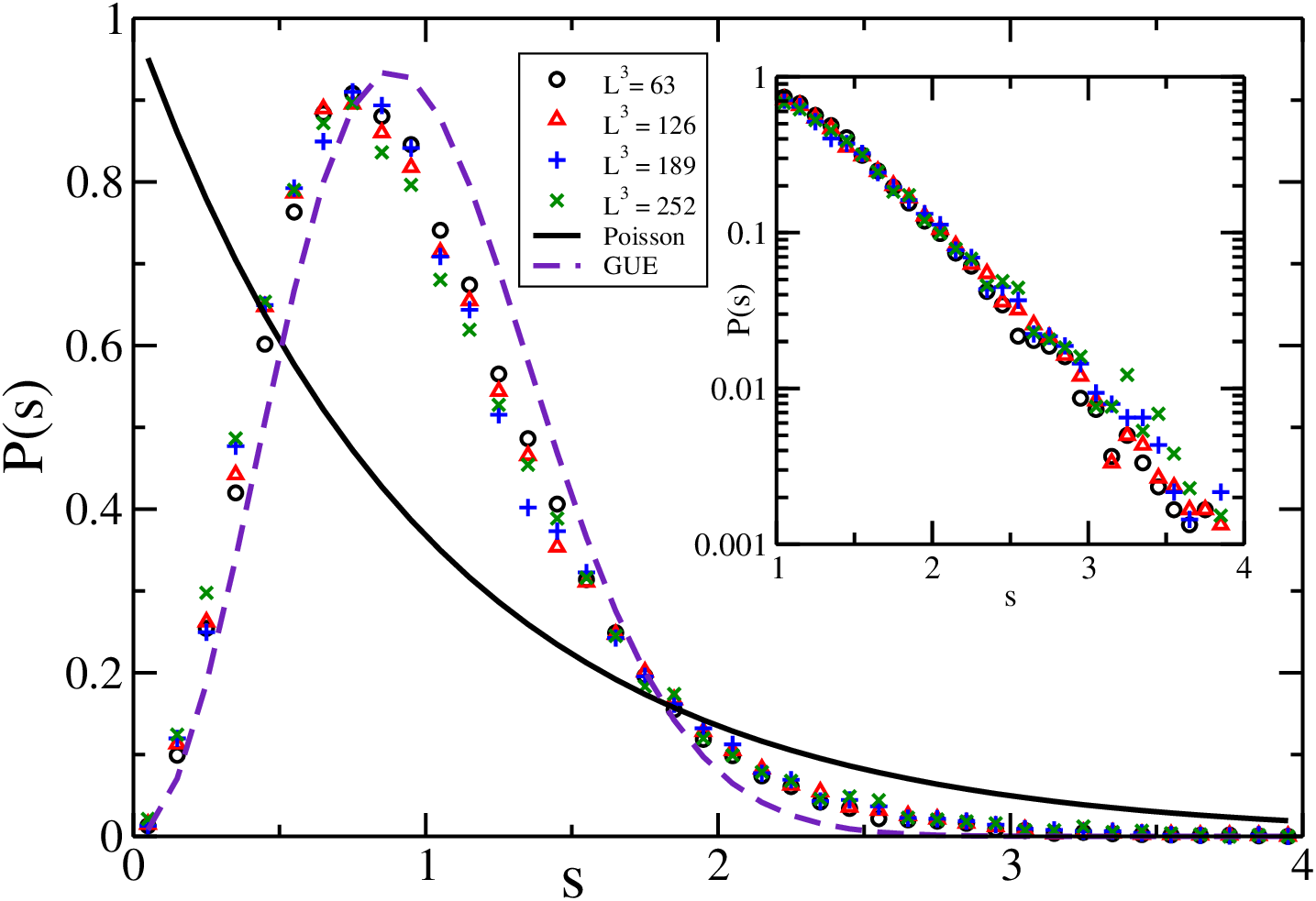}
%   \vspace{3mm}
   \caption{Level spacing distribution $P(s)$ at the origin for the ILM with 2
            massless quarks at the critical temperature $T_c=120$ MeV for
            different spatial volumes.
            The scale invariance indicates the system is undergoing an AT.}
   \label{psnf2m0t120}
  \end{center}
 \end{minipage}
 \hfill
\end{figure}

In figure \ref{cciprnf2m0} the chiral condensate
versus temperature is shown for a range
of system sizes.  We see that the condensate
decreases as the temperature is increased in a manner consistent
with a second order phase transition.
Although we would need to go to much larger volumes to determine the transition
temperature more accurately there is a clear bend in the condensate
around a temperature of $T_c \sim 120$ MeV 
which indicates that the chiral phase transition should occur
 around this region.
Remarkably, as shown in figures \ref{iprnf2m0} and \ref{cciprnf2m0}, 
this is also around the temperature at which the IPR of
the lowest eigenmode begins to rise signaling a transition to localization. 
This finding suggests that both transitions could be
intimately related. Indeed it shows that the phenomenon of Anderson localization 
may be the fundamental mechanism leading to the chiral phase transition in QCD.
In order to further confirm that this novel relation between QCD and 
 the theory of disordered system is not a peculiarity of 
the ILM utilized in this paper 
we are currently carrying out similar numerical calculations in the framework 
 of lattice gauge theory \cite{newaj}. We stress that the 
relation between an Anderson transition and the chiral phase transition is the main result  of this paper.

We now turn to the level statistics.
In order to locate the localization transition temperature more precisely we
look at the number variance $\Sigma^2(\ell)$ near the origin.
In figure \ref{nvonf2m0} we plot this for different temperatures and
spatial volumes.
At $T=100$ MeV, the number variance decreases as the system size increase 
thus indicating that the states are still extended
at this temperature.
At $T=120$ MeV there is still a big difference between the smallest volume
($L^3=63$ fm$^3$) and the larger ones, but the two larger volumes fall
directly on top of each other.
We believe that this is very near the localization transition and that
the smaller volume is just too small to reliably see the proper scaling.
For higher temperatures $T \sim 150$ MeV the number variance moves rapidly
towards the prediction of Poisson statistics $\Sigma^2(\ell)=\ell$.
typical of an insulator.

We observe a similar picture for short range spectral correlators
 such as $P(s)$.
For a fixed volume the spectral correlations 
diminish as we increase the temperature. Thus 
the level spacing distribution
moves (see figure \ref{psnf2m0}) from 
the RMT result (GUE) typical of a metal to Poisson statistics typical 
 of an uncorrelated spectrum.  
By fixing the temperature at $120$ MeV and looking at the spatial volume
dependence (figure \ref{psnf2m0t120}) of $P(s)$,
we observe very little difference for the range of 
sizes considered in agreement with the results for
the number variance.
We also see that there is still level repulsion but the tail of $P(s)$
is exponential as for an insulator.
All these features are typical signatures of an AT.

\begin{table}[b]
 \begin{tabular}{|c|cccc|}
 \hline
 $q$   & 2       & 3       & 4       & 5 \\
 \hline
 ~$D_q$~ & ~2.67(4)~ & ~2.34(5)~ & ~2.08(6)~ & ~1.87(7)~ \\
 \hline
 \end{tabular}
 \caption{Fractal dimensions ($D_q$) for the ILM with 2 massless flavors
          at $T=120$ MeV.}
 \label{dq}
\end{table}

The AT is also 
characterized by the multifractality
of the eigenstates.
We investigate this feature
 by looking at the scaling of $P_q \sim L^{-D_q(q-1)}$ for low eigenvalues
 at $T=120$ MeV for system sizes ranging from $L^3=$ 63--252 fm$^3$.
The resulting fractal dimensions $D_q$ are shown in table \ref{dq}
for an eigenvalue of $\langle\lambda\rangle=0.076$
 which is the smallest eigenvalue for
the $L^3=63$ data.  The IPRs for the remaining sizes were interpolated to
this eigenvalue (see figure \ref{iprnf2m0}).
We clearly see $D_q$ change with $q$
thus showing that the eigenstates are truly multifractal. It is worth 
mentioning that the numerical value of $D_q$ is very sensitive to
small temperature changes.
Thus at $T=111$ MeV we find that $D_2=3.07(5)$ like for a metal ($D_2=3$)
but at $T=150$ MeV,
 $D_2=0.96(4)$ which is closer to the value for an insulator ($D_2=0$).

Just in passing we mention that
 a mobility edge and the corresponding AT has also been
found in the bulk of the spectrum.
As in the quenched case, the temperature at which it appears 
in the central part of the spectrum is higher than the one close to the origin.
This is consistent with the picture that, at a fixed temperature, 
eigenvectors close to the origin 
are more localized than those in the bulk of the spectrum.
Indeed similar results have also been obtained in lattice calculations
 \cite{damgard,lattice}.
The eigenvalue and eigenvector statistics are also similar to the results
shown for the quenched case at the AT so we will not present
them here.

\begin{figure}[t]
 \hfill
 \begin{minipage}[t]{.48\textwidth}
  \begin{center}
   \includegraphics[width=\columnwidth,clip,angle=0]{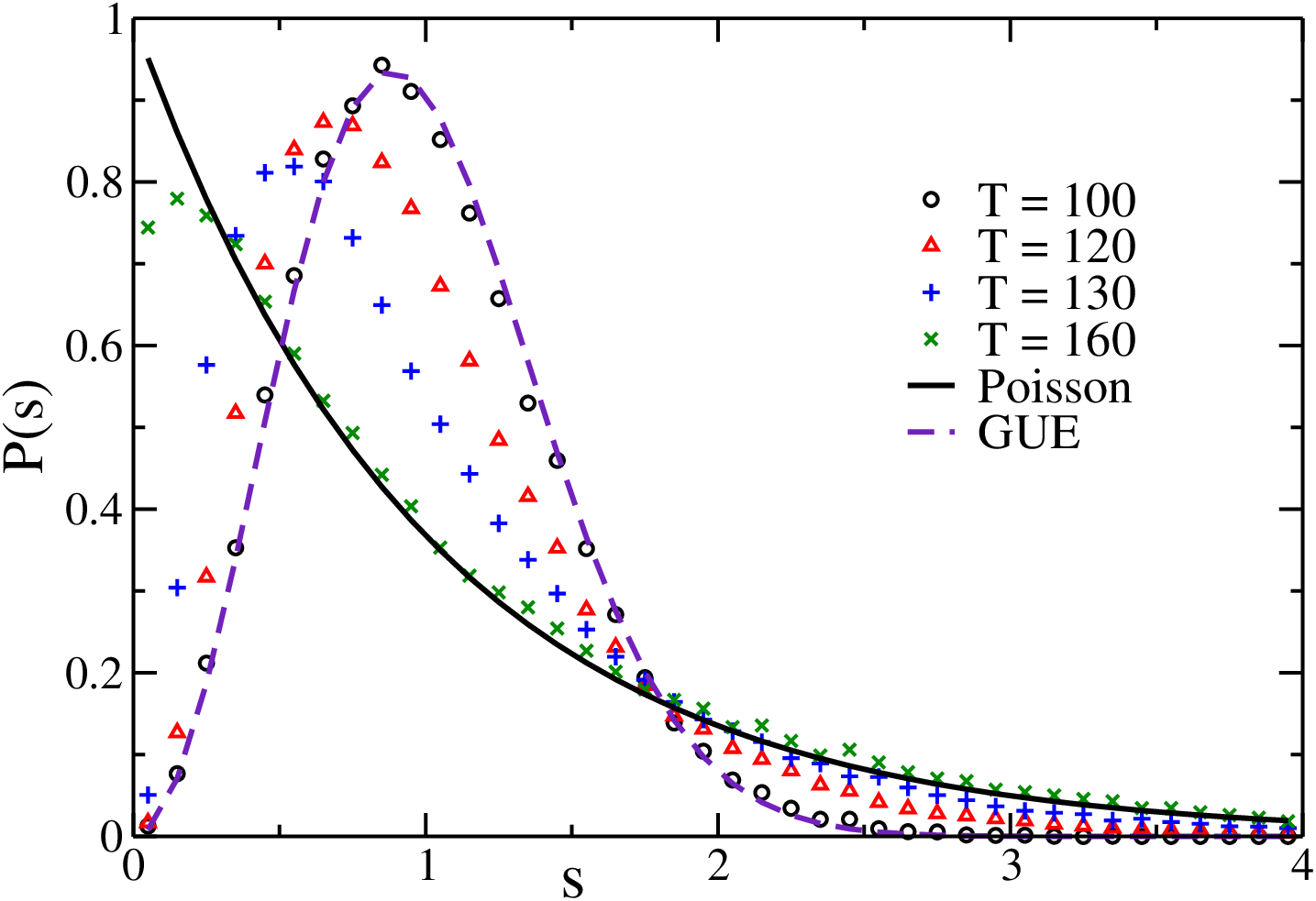}
%   \vspace{3mm}
   \caption{Level spacing distribution $P(s)$ at the origin for the ILM with $N=300$ instantons 
     and 3 massive flavors at different temperatures. As in previous cases a transition to localization 
 is observed as the temperature is increased.}
   \label{psnf3}
  \end{center}
 \end{minipage}
 \hfill
 \begin{minipage}[t]{.48\textwidth}
  \begin{center}
   \includegraphics[width=\columnwidth,clip,angle=0]{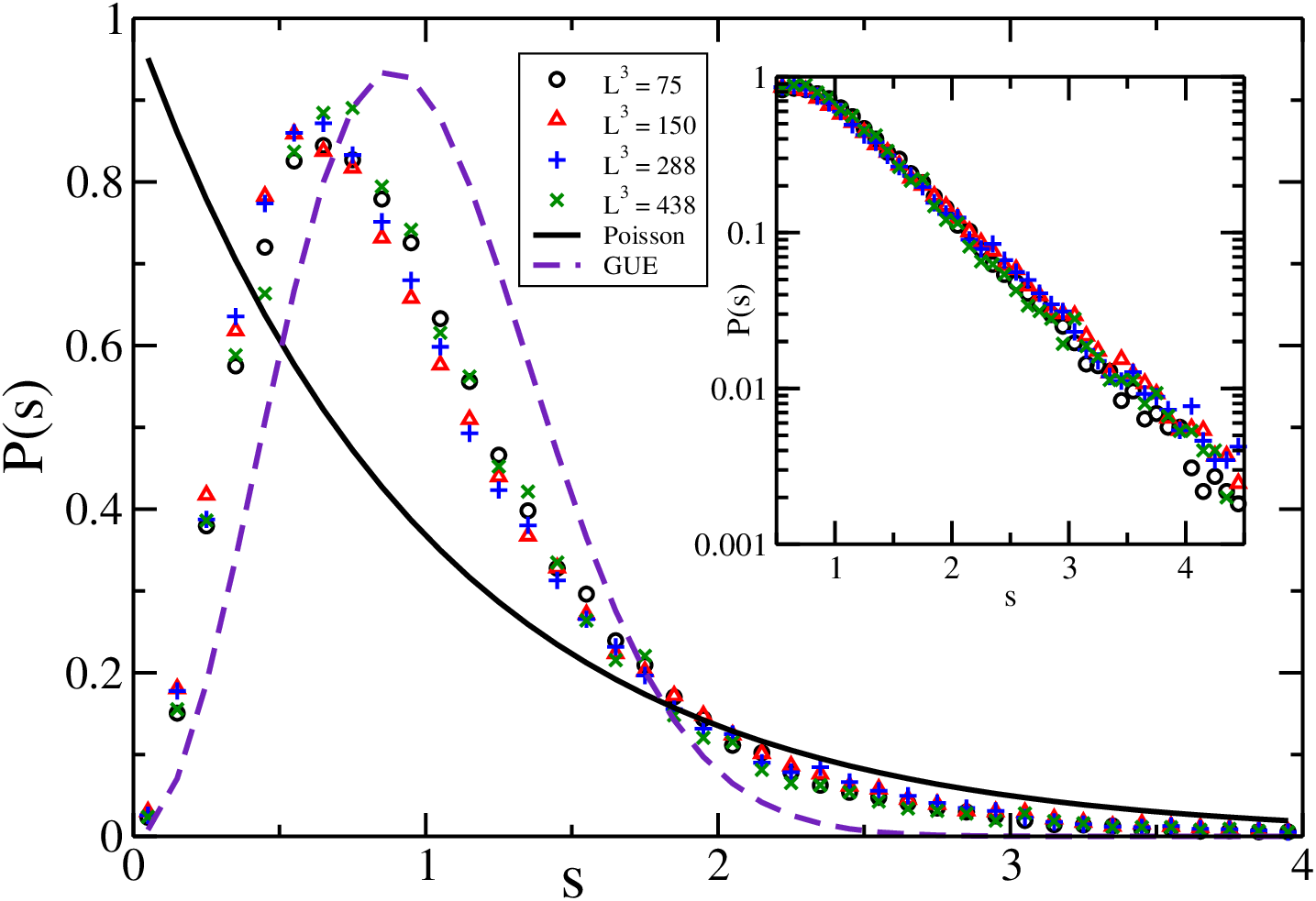}
%   \vspace{3mm}
   \caption{Level spacing distribution $P(s)$ at the origin for the ILM with 3 massive
    flavors at $T=125$ MeV for different spatial volumes. The spectrum is nearly
    scale invariant as expected 
     at an AT.}
   \label{psnf3t125}
  \end{center}
 \end{minipage}
 \hfill
\end{figure}

Finally we discuss the
case of the $N_f=3$ ILM with masses $m_u=m_d=20$ MeV and $m_s=140$ MeV.
As a general comment we observe
 strong similarities with the 
quenched and $N_f=2$ cases studied previously.
Thus, as the plot of the level spacing distribution $P(s)$ in figure \ref{psnf3} shows, 
the lowest eigenmodes
move from extended (metal) to localized (insulator) as the temperature is increased.

More importantly, we have also found that the   
the localization transition of the low-lying eigenmodes of the QCD Dirac 
 operator occurs around the same temperature $T_c \sim 125$ MeV
as the crossover to the chirally symmetric phase
 (which is also not far from the value for two massless flavors). 
However we remark that
 since the lightest masses were still fairly large the eigenvalue
density increases at the origin like in the quenched case
and in lattice simulations with overlap fermions \cite{kiskis}.
Even though the chiral condensate only signals a crossover 
we still see evidence of critical statistics in the critical region
 $T_c \sim 125$ MeV.
Thus, as shown in figure \ref{psnf3t125}, the level spacing at the origin
is scale invariant and exhibits level
repulsion at small distances while having an asymptotically exponential tail.

\section{Discussion and relation to lattice results
\label{secDiscuss}}

In this section we investigate to what extent the
relation between Anderson transition 
 and chiral phase transition reported in previous sections 
 is a particularity of the ILM or could be a real feature
 of the strong interactions.
Obviously a conclusive answer to this question can only 
 be given after a full analysis \cite{newaj} of the 
QCD Dirac operator around the restoration temperature is carried out
 in a direct simulation of QCD on a lattice.

However it is instructive to clearly establish the reasons for the
  appearance of the AT in the ILM.
As is known, 
the fermionic zero modes in the field of an instanton at nonzero temperature
 have an exponential tail $e^{-rT}$ in the spatial directions and are
 oscillatory in the time direction \cite{SS97}.
This suggests that the overlap among different zero modes is essentially
 restricted to nearest neighbors in the spatial directions.
However, in the time direction, different zero modes strongly overlap
 due to the oscillatory character of the eigenmodes.
This situation strongly resembles a 4D disordered conductor 
 in the tight binding approximation (only nearest neighbor hopping) 
 with one dimension (time) much smaller than the rest so the system can be
 considered effectively three dimensional.
It is well established that such a system may undergo an AT depending on the disorder
 strength.
In our case the disorder role is played by temperature since the low-lying eigenvectors
of the QCD Dirac operator decay as $\sim e^{-rT}$.  From this discussion it is natural
 to find an Anderson transition in the
  ILM for a particular value of the temperature.
The principal ingredient to reach the AT is an exponential decay
 of the eigenmodes explicitly depending on the temperature together with the 
 possibility to tune the effective range of the exponential in order to 
 reach the transition region.
Any theory with these features very likely 
 will undergo an AT for some value of the parameters even 
if the objects responsible for localization are not the classical
 instantons. In fact this is not the only scenario 
in which an AT could arise. As mentioned in the introduction, an AT is also expected  
in the case that at a certain 
 temperature the dominant gauge configurations
 are such that the resulting low lying eigenmodes are 
power-law localized $|\psi(r)| \sim 1/r^\alpha$ with $\alpha \sim 3$.
 
We also remark that the AT in the ILM cannot be induced 
by instanton--anti-instanton molecules. As the temperature 
 is increased  overlapping among different instantons becomes more rare and 
 eventually, in the the high temperature limit, the ILM is composed of independent
 instanton--anti-instanton molecules. 
However such situation does not occur at temperatures around the 
 chiral phase transition due to the multifractal character of the eigenstates.
 In the molecular phase the zero modes are attached to a unique
 instanton--anti-instanton pair,
the wavefunction will be localized in between the two pseudoparticles and consequently
the eigenstates cannot be truly multifractal. At most $D_q$ may be a constant not depending on the 
 moment $q$.  Another indication that the molecular phase has not been yet reached at the 
 restoration temperature is the fact that the chiral condensate decreases smoothly over a quite  
broad range of temperatures. If the molecular phase is dominant a gap appears 
 in the spectral density and the condensate vanishes.

We now discuss the relation of our findings with previous
lattice results.
Lattice investigations of the QCD Dirac operator at finite temperature
have attracted considerable attention in recent years. Results close
to the restoration temperature qualitatively agree with the ones
reported in this paper.
It has been observed \cite{damgard,lattice} that eigenvectors
close to the origin are more localized than those of the bulk
(see figures \ref{nf0ipr} and \ref{iprnf2m0}).
The eigenvalue density of the Dirac operator in the quenched ILM more
closely resembles that of lattice simulations of overlap fermions \cite{kiskis}
than those of staggered fermions \cite{damgard}.
This is not surprising since the ILM is based on the low energy topological
modes that overlap fermions are supposed to treat accurately while
staggered fermions (at large lattice spacing) do not.
For two massless flavors the ILM does not have a large eigenvalue density 
near the origin and we can clearly see the decrease in the condensate
as the localization transition occurs.
We cannot make a direct comparison to lattice simulations here due to
the large cost of simulating massless quarks on the lattice.
We also note that a mobility edge has been observed in the low energy
modes of Wilson fermions \cite{GS} however they do not have a chiral symmetry
and we cannot make a direct comparison.
 In order to further clarify this issue, an updated
detailed lattice analysis of level statistics and eigenvectors in the
framework of lattice gauge theory would be highly desirable.

Finally we mention that the true nature of the topological objects responsible
for \SB{} and confinement has not been fully settled in lattice QCD.
Although there is some support for instanton-like objects there are other
studies that suggest a different picture.
Recently it has been argued that the topological charge may not be localized
 in four dimensional regions of space-time but rather spread out over
 a lower dimensional manifold \cite{lattop}.
The resulting low-lying eigenmodes of the QCD Dirac operator are still
 delocalized and fractal as we observe in the ILM.
This is also not far from the results of the ILM in \cite{aj,VO} where it was
 found that the low-lying eigenmodes are delocalized in the sense that in the
 thermodynamic limit the level statistics 
 are described by random matrix theory.
It would be interesting to compare these predictions with results from
 overlap fermions at zero temperature.

\section{Conclusions\label{secConcl}}

We have studied the localization properties of the eigenvalues and
eigenvectors of the QCD Dirac operator in a instanton liquid model
at finite temperature.
Due to the chiral symmetry of the model, eigenstates close to the
origin are more localized than those near the center for any temperature.
Near the origin we found a clear transition from 
 delocalized to localized states as the
temperature is increased.
There is a also mobility edge in  the bulk of the spectrum that moves toward 
 the region of lower energies as the temperature is increased.
In both cases, around this mobility edge, the eigenvectors are multifractal and
spectral correlations are well described by critical statistics as
for a disordered system undergoing an AT.
Remarkably both the transition to localization close to the origin
and the chiral phase transition signaling  the restoration of the chiral
symmetry occur at roughly the same temperature.
This suggests that the phenomenon of Anderson localization may be
 the fundamental mechanism driving the chiral phase transition.

\begin{acknowledgments}
AMG thanks Jac Verbaarschot for iluminating discussions.
AMG was supported by a Marie Curie Outgoing Fellowship,
 contract MOIF-CT-2005-007300.
JCO was supported in part by U.S. DOE grant DE-FC02-01ER41180.
\end{acknowledgments}


\begin{thebibliography}{9}
%\vspace{-15mm}

\bibitem{lattrans}
 F. Karsch, E. Laermann and A. Peikert,
 \emph{Nucl. Phys.} {\bf B605} (2001) 579;
 C. Gattringer, P.E.L. Rakow, A. Schafer and W. Soldner,
 \emph{Phys. Rev.} D {\bf 66} (2002) 054502.

\bibitem{anderson} P.W. Anderson,
 \emph{Phys. Rev.} {\bf 109} (1958) 1492.

\bibitem{polyakov} A. Belavin, A. Polyakov, A. Schwartz and Y. Tyupkin,
 \emph{Phys. Lett.} {\bf 59} (1975) 85;
 G. 't Hooft, \emph{Phys. Rev. Lett.} {\bf 37} (1976) 8.

\bibitem{diakonov} D. Diakonov and V. Petrov,
 \emph{Nucl. Phys.} {\bf B245} (1984) 259.

\bibitem{shuryak} E. Shuryak,
 \emph{Nucl. Phys.} {\bf B203} (1982) 93,116,140.

\bibitem{SS97} T. Sch\"afer and E. Shuryak,
 \emph{Rev. Mod. Phys.} {\bf 70} (1998) 323.

\bibitem{SSV} T. Shafer, E. Shuryak and J.J.M. Verbaarschot,
 \emph{Nucl. Phys.} {\bf B412} (1994) 143.

\bibitem{latins} M.-C. Chu, et. al.,
 \emph{Phys. Rev.} D {\bf 49} (1994) 6039;
 C. Michael and P.S. Spencer, \emph{Phys. Rev.} D {\bf 52} (1995) 4691.

\bibitem{bank} T. Banks and A. Casher,
 \emph{Nucl. Phys.} {\bf B169} (1980) 103.

\bibitem{diakonov1} D. Diakonov and P. Petrov,
 \emph{Phys. Lett.} {\bf B147} (1984) 351;
 \emph{Nucl. Phys.} {\bf B272} (1986) 457;
 hep-ph/9602375.

\bibitem{smilga} A.V. Smilga,
 \emph{Phys. Rev.} D {\bf 46} (1992) 5598.

\bibitem{zahed} R.A. Janik, M.A. Nowak, G. Papp and I. Zahed,
 \emph{Phys. Rev. Lett.} {\bf 81} 264 (1998).

\bibitem{VO} J.C. Osborn and J.J.M. Verbaarschot,
 \emph{Phys. Rev. Lett.} {\bf 81} 268 (1998);
 \emph{Nucl. Phys.} {\bf B525} 738 (1998).

\bibitem{levitov90}
 L.S. Levitov, \emph{Phys. Rev. Lett.} {\bf 64} (1990) 547;
 A.D. Mirlin, Y.V. Fyodorov, F.-M. Dittes, J. Quezada and T.H. Seligman,
 \emph{Phys. Rev.} E {\bf 54} (1996) 3221;
 E. Cuevas, \emph{Phys. Rev.} B {\bf 68} (2003) 024206,184206.

\bibitem{par} A. Parshin and H.R. Schober,
 \emph{Phys. Rev.} B {\bf 57} (1998) 10232.

\bibitem{kravtsov97} V.E. Kravtsov and K.A. Muttalib,
 \emph{Phys. Rev. Lett.} {\bf 79} (1997) 1913;
 S. Nishigaki, \emph{Phys. Rev.} E {\bf 59} (1999) 2853;
 F. Evers and A.D. Mirlin,
 \emph{Phys. Rev. Lett.} {\bf 84} (2000) 3690;
 E. Cuevas, M. Ortu\~no, V. Gasparian and A. P\'erez-Garrido,
 \emph{Phys. Rev. Lett.} {\bf 88} (2002) 016401.

\bibitem{janssen} M. Janssen, \emph{Phys. Rep.} {\bf 295} (1998) 1. 

\bibitem{ono} G. Yeung and Y. Oono,
 \emph{Europhys. Lett.} {\bf 4} (1987) 1061.

\bibitem{aj} A.M. Garcia-Garcia and J.C. Osborn,
 \emph{Phys. Rev. Lett.} {\bf 93} (2004) 132002.

\bibitem{lattice}
 M. Garcia Perez, A. Gonzalez-Arroyo, A. Montero and P. van Baal,
 \emph{JHEP} 9906 (1999) 001;
 M. G\"ockeler, P.E.L. Rakow, A. Sch\"afer, W. S\"oldner and T. Wettig,
 \emph{Phys. Rev. Lett.} 87 (2001) 042001;
 C. Gattringer, M. G\"ockeler, P.E.L. Rakow, S. Schaefer and A. Sch\"afer,
 \emph{Nucl. Phys.} {\bf B618} (2001) 205;
 C. Gattringer and S. Schaefer,
 \emph{Nucl. Phys.} {\bf B654} (2003) 30;
 B. Lucini, M. Teper and U. Wenger,
 \emph{Nucl. Phys.} {\bf B715} (2005) 461;
 E.-M. Ilgenfritz, M. M\"uller-Preussker and D. Peschka,
 \emph{Phys. Rev.} D {\bf 71} (2005) 116003.

\bibitem{mehta} M.L. Mehta,
 {\it Random Matrices}, Academic Press, New York (1991) 2nd edition.

\bibitem{wegner} F. Wegner, {\emph Z. Phys.} B {\bf 36} (1980) 209;
 B.L. Altshuler, V.E. Kravtsov and I.V. Lerner,
 in: {\emph Mesoscopic Phenomena in Solids} edited by
 B.L. Altshuler, P.A. Lee and R.A. Webb, North Holland, Amsterdam, (1991).
 V.I. Falko and K.B. Efetov, {\emph Europhys. Lett.} {\bf32} (1995) 627;

\bibitem{mirlin} A. Mirlin, \emph{Phys. Rep.} {\bf 326} (2000) 259.

\bibitem{sko} B.I. Shklovskii, B. Shapiro, B.R. Sears, P. Lambrianides and
 H.B. Shore, \emph{Phys. Rev.} B {\bf 47} (1993) 11487.

\bibitem{chi} B.L. Altshuler, I.K. Zharekeshev, S.A. Kotochigova and 
 B.I. Shklovskii, \emph{JETP} 67 (1988) 62.

\bibitem{nikolic} B. Nikolic,
 \emph{Phys. Rev.} B {\bf 64} (2001) 014203.

\bibitem{evangelou} S. Xiong and S.N. Evangelou,
 \emph{Phys. Rev.} B {\bf 64} (2001) 113107.

\bibitem{verges}
 C. Mudry, P.W. Brouwer and A. Furusaki,
 \emph{Phys. Rev.} B {\bf 62} (2000) 8249;
 G. Chiappe, E. Louis, M.J. Sanchez and J.A. Verges,
 \emph{Phys. Rev.} B {\bf 69} (2004) 201405.

\bibitem{kiskis} J. Kiskis and R. Narayanan,
 \emph{Phys.Rev.} D {\bf 64} (2001) 117502.

\bibitem{chen} K.A. Muttalib, Y. Chen, M.E.H. Ismail and V.N. Nicopoulos,
 \emph{Phys. Rev. Lett.} {\bf 71} (1993) 471;
 Y. Chen and K.A. Muttalib,
 \emph{J. Phys.:Condens. Matter} {\bf 6} (1994) L293.

\bibitem{ant3} A.M. Garcia-Garcia and J.J.M. Verbaarshot,
 \emph{Phys. Rev.} E {\bf 67} (2003) 046104.

\bibitem{kazu} A.M. Garcia-Garcia and K. Takahashi,
 \emph{Nucl.Phys.} {\bf B700} (2004) 361.

\bibitem{tsv} V.E. Kravtsov and A.M. Tsvelik,
 \emph{Phys. Rev.} B {\bf 68} (2000) 9888. 

\bibitem{calo} B. Sutherland,
 \emph{J. Math. Phys.} {\bf 12} (1971) 246;
 F. Calogero, \emph{J. Math. Phys.} {\bf 10} (1969) 2191.

\bibitem{damgard} P.H. Damgaard, U.M. Heller, R. Niclasen and K. Rummukainen,
 \emph{Nucl.Phys.} {\bf B583} (2000) 347.

\bibitem{SS96} T. Schaefer and E.V. Shuryak,
 \emph{Phys. Rev.} D {\bf 53} (1996) 6522.

\bibitem{newaj} A.M. Garcia-Garcia and J.C. Osborn, in prepartion.

\bibitem{GS} M. Golterman and Y. Shamir,
 \emph{Phys. Rev.} D {\bf 68} (2003) 074501;
 \emph{Phys. Rev.} D {\bf 71} (2005) 071502;
 \emph{Phys. Rev.} D {\bf 72} (2005) 034501.

\bibitem{lattop}
 I. Horvath, N. Isgur, J. McCune and H.B. Thacker,
 \emph{Phys. Rev.} D {\bf 65} (2002) 014502;
 I. Horvath, et. al.,
 \emph{Phys. Rev.} D {\bf 66} (2002) 034501;
 I. Horvath, \emph{Nucl. Phys.} {\bf B710} (2005) 464;
 Erratum-ibid. {\bf B714} (2005) 175;
 I. Horvath, et.al.,
 \emph{Phys. Lett.} {\bf B612} (2005) 21;
 A. Alexandru, I. Horvath and J.B. Zhang,
 \emph{Phys. Rev.} D {\bf 72} (2005) 034506;
 V. Weinberg, E.-M. Ilgenfritz, K. Koller, Y. Koma, G. Schierholz and
 T. Streuer, \emph{PoS} LAT2005 (2005) 171;
 C. Bernard, et.al., \emph{PoS} LAT2005 (2005) 299.

\end{thebibliography}
\end{document}